\documentclass[5p,twocolumn,number,times]{elsarticle}
\usepackage{amsmath}
\usepackage{amssymb}
\usepackage{graphicx}
\usepackage{hyperref}
\usepackage[english]{babel}
\usepackage[usenames,dvipsnames]{color}

\newcommand{\Tr}{\operatorname{Tr}}
\newcommand{\func}[1]{\operatorname*{#1}}

\journal{Physics Letters A}
\begin{document}

\begin{frontmatter}

\title{Conditional non-Hermitian acceleration of multiphoton atomic transitions}
\author[if]{M.V.S. de Paula}
\author[if]{A.P. Costa}
\author[if,cif]{A.V. Dodonov\corref{cor1}}
\ead{adodonov@unb.br}
\cortext[cor1]{Corresponding author.}
\address[if]{Instituto de F\'{\i}sica, Universidade de Bras\'{\i}lia, Caixa Postal 04455, CEP
70910-900, Bras\'{\i}lia, DF, Brasil}
\address[cif]{International Center of Physics, Institute of Physics, University of Bras\'{\i}lia, 70910-900, Bras\'{\i}lia, DF, Brazil}

\begin{abstract}
{Continuous monitoring can conditionally accelerate otherwise slow multiphoton transitions through state-selective no-jump filtering. We consider a $\Lambda$-type three-level atom and condition the evolution on the absence of photon emission associated with an auxiliary monitored decay channel. The resulting no-jump dynamics is governed by a non-Hermitian Rabi-type Hamiltonian. Using Floquet theory and Brillouin--Wigner perturbation theory, we derive effective two-state descriptions of odd-multiphoton resonances in the semiclassical and quantum Rabi models. We characterize the conditional acceleration by the time required to reach a prescribed target-state population and compare instantaneous and coarse-grained definitions. Numerical results agree with the analytical predictions in their regimes of validity. However, the ideal minimum mean time required to obtain one successful preparation increases because of the decreasing no-jump survival probability. Thus, monitoring accelerates selected multiphoton dynamics without providing an overall preparation-time advantage.}
\end{abstract}


\end{frontmatter}

\section{Introduction}

The semiclassical and quantum Rabi models are paradigmatic descriptions of
the interaction of a two-level system with a classical oscillatory field and
a quantized bosonic mode, respectively. Beyond the ordinary one-photon
resonance, these models support multiphoton resonances when the atomic
transition frequency approaches an integer multiple of the field frequency.
For the standard purely transverse (dipole) interaction, only odd-photon
transitions between the two atomic states are allowed. Multiphoton
resonances and the associated Bloch--Siegert shifts caused by
counter-rotating terms were observed in a classical two-level analogue---an
optical ring resonator with two orthogonal linear polarizations \cite{woerd}%
. In the quantum Rabi model, counter-rotating terms in the Hamiltonian
produce resonant multiphoton atom--field exchange, including three-photon
vacuum Rabi oscillations and adiabatic three-photon conversion in the
large-detuning regime \cite{MaLaw2015,Garziano2015}.

The analysis of strongly driven two-level systems is naturally formulated
within Floquet theory. Shirley mapped the Schr\"odinger equation with a
periodic Hamiltonian onto a stationary quasienergy problem in an extended
Hilbert space \cite{shirley}, and Sambe developed the corresponding
time-independent Hilbert-space representation \cite{Sambe1973}. Floquet
methods have since become standard tools for driven quantum systems,
including dissipative two-state dynamics and strong-driving regimes beyond
the rotating-wave approximation \cite{Grifoni1998}. Duvall \textit{et al.}
developed a nonperturbative analysis specifically for multiphoton excitation
of a driven two-level atom \cite{duvall}. A complementary
resonance-expansion approach was introduced to classify effective
transitions generated beyond the rotating-wave approximation and to
determine the accompanying frequency shifts \cite{Klimov2003}.
Effective-Hamiltonian techniques based on Floquet representations were
subsequently formulated for periodically perturbed quantum-optical systems 
\cite{ad7}.

The effective coupling associated with an odd-multiphoton resonance
decreases rapidly with the resonance order. Higher-order transitions
therefore occur on increasingly long time scales and become more vulnerable
to relaxation, dephasing, and parameter fluctuations. Related selective odd-$%
k$-photon interactions have also been derived in the Rabi--Stark model,
where the additional Stark coupling makes the resonance dependent on the
bosonic occupation and the effective interaction strength decreases rapidly
with $k$ \cite{Cong2020}. Approximate analytical expressions for the
dissipative semiclassical Rabi model near the three-photon resonance were
previously derived and compared with the corresponding quantum Rabi dynamics 
\cite{marinho2}. These results motivate the search for mechanisms that
shorten the multiphoton population-transfer time without modifying the
underlying purely transverse interaction.

A natural route is provided by the quantum-trajectory description of open
systems. When a decay channel is continuously monitored and the evolution is
conditioned on the absence of the associated quantum jump, the unnormalized
conditional state evolves under an effective non-Hermitian Hamiltonian. Its
squared norm, or equivalently the trace of the corresponding unnormalized
density operator, gives the probability of the selected no-jump record \cite%
{Dalibard1992,Molmer1993}. The non-Hermitian generator therefore has a
direct operational meaning: it describes a postselected subset of the
trajectories of a physical open quantum system. Continuous monitoring can
modify the effective spectrum and the conditioned population dynamics, and
can thereby shorten the time required to reach a prescribed target-state
population. Such a conditional acceleration, however, must be assessed
together with the probability of realizing the required no-jump trajectory.

The possibility of reducing characteristic evolution times with
non-Hermitian generators has been discussed extensively in the context of
nonunitary quantum dynamics \cite{Bender2007,Assis2008}. For non-Hermitian
representations of genuinely unitary theories, the physical inner product
restores the usual Hermitian speed constraints \cite{Mostafazadeh2007}.
Dissipative non-Hermitian dynamics is fundamentally different because the
evolution is nonunitary and the norm of the unnormalized state has a direct
probabilistic interpretation. Experimentally, continuously monitored
superconducting circuits have enabled the reconstruction of individual
quantum trajectories \cite{Murch2013} and the characterization of
postselected non-Hermitian dynamics \cite{Naghiloo2019}. Dynamically
controlled non-Hermitian systems have also been used to realize
nonreciprocal state transfer \cite{Abbasi2022}. Related theoretical and
experimental studies have shown that reductions in evolution time can be
accompanied by a decreasing probability of successful postselection \cite%
{Li2023,Yuan2026}. These results emphasize the distinction between faster
dynamics within a conditioned ensemble and the overall efficiency of a
repeated state-preparation protocol.

{In this work, we apply the no-jump conditional-filtering mechanism to odd-multiphoton transitions in the semiclassical and quantum Rabi models. We consider a three-level atom in a \(\Lambda\) configuration, subject to continuous monitoring of the decay from the excited state to an auxiliary state. Conditioning on the absence of this emission generates a non-Hermitian contribution to the Rabi-type Hamiltonian. Using Floquet theory and Brillouin–Wigner projection-operator perturbation theory, we derive effective two-state Hamiltonians near the multiphoton resonances, including the high-order multiphoton couplings, the accompanying frequency shifts, and the leading dispersive correction produced by the weakly coupled auxiliary level. We characterize the conditional dynamics by the instantaneous first-passage time \(T_p\) at which a prescribed target-state population \(p\) is reached, together with the centered coarse-grained time \(\overline{T}_p\), and define the corresponding conditional speedup relative to the Hermitian evolution. For high target populations, the instantaneous criterion can be sensitive to micromotion; however, for certain parameter regimes in which the threshold remains attainable after coarse graining, both \(T_p\) and \(\overline{T}_p\) decrease as the monitored decay rate increases. We also calculate the maximum attainable target population and derive the ideal minimum mean preparation times \(\tau_p\) and \(\overline{\tau}_p\) for repeated no-jump attempts. In contrast to the conditional transfer times, these mean preparation times increase because of the decreasing survival probability. Thus, faster conditioned multiphoton transfer does not imply a faster overall preparation protocol. We test the effective analytical description against full three-level numerical dynamics in both the semiclassical and quantum regimes, and compare the conditioned dynamics with the complete unconditioned master equation.

The generic conditional-filtering mechanism itself is not specific to multiphoton transitions. The main multiphoton-specific contribution of the present work is the explicit derivation and quantitative validation of this mechanism in a regime where coherent transfer is generated only at high perturbative order through counter-rotating virtual transitions. In particular, the effective multiphoton couplings and the accompanying frequency shifts are derived from the underlying Rabi models rather than introduced phenomenologically. The weak auxiliary level is retained through its leading complex dispersive correction, and the threshold dependence, coarse-grained transfer dynamics, survival probability, and joint atom–field target fidelity are tested against the full three-level dynamics. Monitoring therefore does not create a distinct multiphoton coupling; it conditionally filters the slow multiphoton dynamics generated by the Rabi Hamiltonian.
}

The remainder of this paper is organized as follows. Section \ref{sec2}
introduces the monitored three-level model and the conditioned no-jump
dynamics. Section \ref{sec3} develops the semiclassical analysis, including the Floquet
effective Hamiltonian, the dispersive correction induced by the auxiliary
level, the conditional transfer time, the ideal minimum mean time to
successful preparation, and comparison with numerical results. Sections \ref{sec5} and \ref{sec6} address the quantum
Rabi model and compare the effective analytical description with direct
numerical simulations of the full three-level dynamics. Section \ref{sec7}
summarizes the main conclusions.

\section{Monitored three-level atom and conditioned dynamics}

\label{sec2}

\begin{figure}[tbh]
\begin{center}
\includegraphics[width=0.4\textwidth]{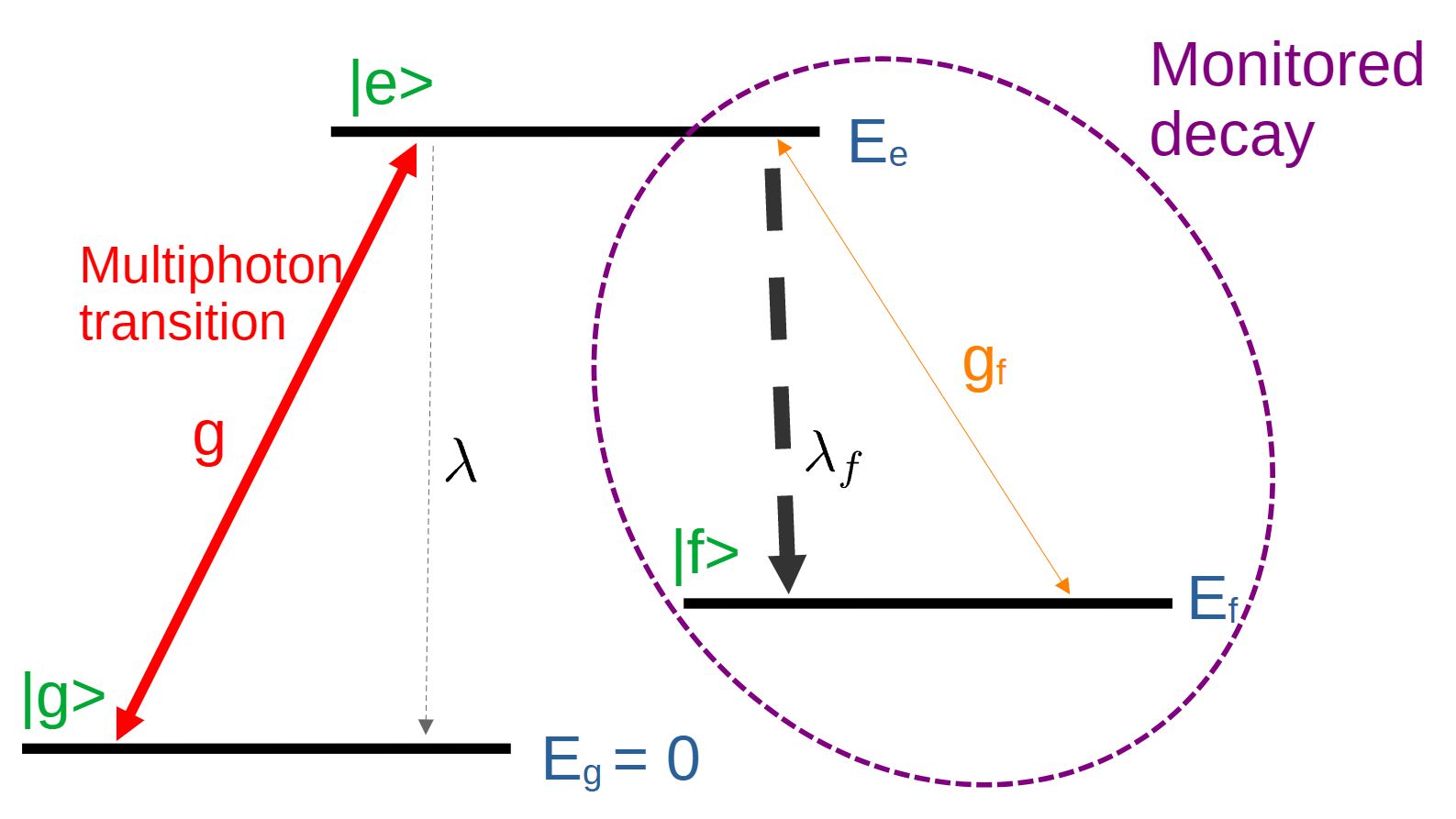} {}
\end{center}
\caption{Level scheme of the monitored $\Lambda$-atom. The resonant
odd-multiphoton transition couples $|g\rangle$ and $|e\rangle$, while the
auxiliary state $|f\rangle$ is weakly and off-resonantly coupled to $%
|e\rangle$ by the classical field and is also connected to it by the
monitored decay $|e\rangle\rightarrow|f\rangle$. The conditioned dynamics is
obtained by postselecting trajectories in which no quantum jump associated
with this incoherent decay channel occurs.}
\label{fig1}
\end{figure}

We begin with a Lambda-type three-level atom driven by a monochromatic
classical field. Its Hamiltonian is%
\begin{eqnarray}
H &=&E_{g}\sigma _{gg}+E_{e}\sigma _{ee}+E_{f}\sigma _{ff}+2g\cos \left(
\omega t\right) \left( \sigma _{eg}+\sigma _{ge}\right)  \notag \\
&&+2g_{f}\cos \left( \omega t\right) \left( \sigma _{ef}+\sigma _{fe}\right)
\,.
\end{eqnarray}%
Here $|e\rangle $ is the excited state, whereas $|g\rangle $ and $|f\rangle $
are two nondegenerate lower states, and $\sigma _{kl}=|k\rangle \langle l|$.
The level scheme is shown in Fig.~\ref{fig1}, where $\omega$ is the field
frequency, $E_{k}$ are the atomic energy levels (we set $E_{g}=0$ throughout), and $g$ and $g_{f}$ are the
coupling constants for the two dipole-allowed transitions. We assume $g_{f}\ll\left|E_{e}-E_{f}-\omega\right|$, and incorporate
the leading effect of this off-resonant coupling
perturbatively through the complex dispersive shift in the effective two-level model, as shown in ~\ref{ap3}.

In the weak-coupling Markovian regime, the unconditioned density operator $%
\rho $ obeys the Gorini-Kossakowski-Sudarshan-Lindblad (GKSL) master
equation \cite{Gorini1976,Lindblad1976,Hebert}. Setting $\hbar =1$,%
\begin{equation}
\dot{\rho}=-i\left[ H,\rho \right] +\frac{\lambda }{2}\mathcal{D}\left(
\sigma _{ge}\right) \rho +\frac{\gamma _{\phi }}{2}\mathcal{D}\left( \sigma
_{z}\right) \rho +\frac{\lambda _{f}}{2}\mathcal{D}\left( \sigma
_{fe}\right) \rho \,,  \label{cme}
\end{equation}%
where $\lambda $ and $\lambda _{f}$ are the relaxation rates for the
transitions $|e\rangle \rightarrow |g\rangle $ and $|e\rangle \rightarrow
|f\rangle $, respectively. The parameter $\gamma _{\phi }$ is the
pure-dephasing rate of the $|e\rangle $--$|g\rangle $ transition, and $%
\sigma _{z}=|e\rangle \langle e|-|g\rangle \langle g|$. The standard
Lindblad dissipator is%
\begin{equation}
\mathcal{D}\left( X\right) \rho \equiv 2X\rho X^{\dagger }-X^{\dagger }X\rho
-\rho X^{\dagger }X\,.
\end{equation}

We assume that the atom is initially prepared in $|e\rangle$ and that the
incoherent radiative transition $|e\rangle \rightarrow |f\rangle$,
accompanied by the emission of a photon, is continuously monitored. The
conditioned ensemble is selected by requiring that no photon of frequency $%
E_{e}-E_{f}$ be detected during the interval $[0,t)$. In the quantum-jump
representation \cite{Dalibard1992,Molmer1993}, this is the subensemble of
trajectories containing no $|e\rangle \rightarrow |f\rangle $ jump. The
associated unnormalized conditional density operator $\rho_c$ obeys%
\begin{equation}
\dot{\rho}_{c}=-i\left[ H_{c}\rho _{c}-\rho _{c}H_{c}^{\dagger }\right] +%
\frac{\lambda }{2}\mathcal{D}\left( \sigma _{ge}\right) \rho _{c}+\frac{%
\gamma _{\phi }}{2}\mathcal{D}\left( \sigma _{z}\right) \rho _{c}\,,
\label{me}
\end{equation}%
where%
\begin{equation}
H_{c}=H-i\kappa \sigma _{ee}~,~\kappa \equiv \frac{\lambda _{f}}{2}\,.
\label{Hc}
\end{equation}%
The probability of the selected no-jump record up to time $t$ is $P^{(c)}(t)=%
\Tr[\rho_{c}(t)]$, whereas the conditional expectation value of an
observable $X$ is $\left\langle X\right\rangle =\Tr%
[X\rho_{c}(t)]/P^{(c)}(t)$.

We show below that the conditioned transition $|e\rangle \rightarrow
|g\rangle$ is accelerated near the odd-multiphoton resonance $%
\Omega=E_{e}-E_{g}\approx(2K+1)\omega$, with $K=1,2,\ldots$. We characterize
this acceleration by the first time required to reach a prescribed
conditional population of the target-state. This transfer time decreases as the
monitored decay rate is increased over the range in which the desired target
population remains attainable. Since the effect is obtained by
postselection, the reduction of the conditional transfer time is accompanied
by a decreasing no-jump probability. Consequently, the conditional
acceleration must be distinguished from the mean time required to obtain one
successful preparation under repeated attempts.

\section{Conditioned semiclassical Rabi dynamics}

\label{sec3}

The conditioned coherent dynamics is generated by the non-Hermitian
semiclassical Rabi Hamiltonian%
\begin{equation}
H_{c}(t)=\frac{\Omega }{2}\sigma _{z}+2g\cos \left( \omega t\right) \left(
\sigma _{+}+\sigma _{-}\right) -i\kappa |e\rangle \langle e|~,
\end{equation}%
where $\sigma _{+}=|e\rangle \langle g|$ and $\sigma _{-}=\sigma
_{+}^{\dagger }$. In our analytical approach we use $\Omega\rightarrow \Omega+\func{Re}%
\left( \delta _{e}\right) $ and $\kappa \rightarrow \kappa -\func{Im}\left(
\delta _{e}\right) $, where the complex dispersive shift $\delta _{e}$ due
to the off-resonant level $|f\rangle $ is calculated in \ref{ap3}. Moving to
a rotating frame at frequency $\omega $, we separate the interaction-picture
Hamiltonian into a time-independent part and a counter-rotating perturbation,%
\begin{equation}
H_{I}(t)=H_{0}+V(t)
\end{equation}%
with%
\begin{equation}
H_{0}=-\frac{\Delta }{2}\sigma _{z}-i\kappa |e\rangle \langle e|+g(\sigma
_{+}+\sigma _{-})
\end{equation}%
\begin{equation}
V(t)=g\left( e^{2i\omega t}\sigma _{+}+e^{-2i\omega t}\sigma _{-}\right) \,.
\end{equation}%
Here $\Delta =\omega -\Omega $ is the one-photon detuning. We work in the
perturbative regime $\left\vert g\right\vert \ll \left\vert \Delta
\right\vert $, which near the odd-multiphoton resonances implies $g\ll
2\omega $. The analytical treatment also assumes that $\lambda $ and $\gamma
_{\phi }$ are small compared with the relevant effective multiphoton
coupling strengths, denoted below by $M_{0K}$. The complete conditioned
master equation (\ref{me}) is retained for the numerical comparisons.

In the basis $\{|g\rangle ,|e\rangle \}$, the right eigenvalues and
eigenvectors of the non-Hermitian unperturbed Hamiltonian are defined by $%
H_{0}|\phi _{\pm }^{R}\rangle =\lambda _{\pm }|\phi _{\pm }^{R}\rangle $ and
read%
\begin{equation}
\lambda _{\pm }=-\frac{i\kappa }{2}\pm \chi
\end{equation}%
\begin{equation}
|\phi _{\pm }^{R}\rangle =\frac{g|g\rangle +(\pm \chi -\delta )|e\rangle }{%
\mathcal{N}_{\pm }}~,\,
\end{equation}%
where%
\begin{equation}
\delta =\frac{\Delta +i\kappa }{2}~,~\chi =\sqrt{g^{2}+\delta ^{2}}~,~%
\mathcal{N}_{\pm }=\sqrt{2\chi (\chi \mp \delta )}~.
\end{equation}%
For $\Delta <0$ and $\left\vert \Delta \right\vert \gg g\gg \kappa $, the
dressed states approach the bare states, $|\phi _{+}^{R}\rangle \approx
|e\rangle $ and $|\phi _{-}^{R}\rangle \approx |g\rangle $, while $\func{Im}%
\left( \chi \right) \approx -\kappa /2$. This limit makes the effective
dressed-state transfer equivalent to the desired atomic transition.

Because the effective Hamiltonian is non-Hermitian, the projection formalism
requires the corresponding left eigenvectors, defined by $\langle \phi _{\pm
}^{L}|H_{0}=\lambda _{\pm }\langle \phi _{\pm }^{L}|$:%
\begin{equation}
\langle \phi _{\pm }^{L}|=\frac{g\langle g|+(\pm \chi -\delta )\langle e|}{%
\mathcal{N}_{\pm }}~.\,
\end{equation}%
With the chosen normalization, the left and right eigenvectors satisfy the
biorthogonality and completeness relations $\langle \phi _{\alpha }^{L}|\phi
_{\beta }^{R}\rangle =\delta _{\alpha \beta }\,$ and $\sum_{i}|\phi
_{i}^{R}\rangle \langle \phi _{i}^{L}|=\hat{I}$, respectively. These
relations are used throughout the effective-Hamiltonian construction.

\subsection{Effective Hamiltonian\label{sec4}}

The interaction-picture Hamiltonian $H_{I}\left( t\right) $ is periodic, $%
H_{I}(t+T)=H_{I}(t)\,$, with period $T=\pi /\omega $. Floquet theory then
permits solutions of the Schr\"{o}dinger equation to be written as $|\varphi
\left( t\right) \rangle =e^{-i\varepsilon t}|u(t)\rangle \,$, where $%
|u(t+T)\rangle =|u(t)\rangle \,$ and $\varepsilon $ is a quasienergy \cite%
{shirley,Sambe1973,Grifoni1998}. We expand the interaction-picture
wavefunction in the biorthogonal dressed-state basis and in Fourier
harmonics,%
\begin{equation}
|\psi \left( t\right) \rangle =\sum_{\alpha =\pm }\sum_{n=-\infty }^{\infty
}W_{\alpha ,n}\left( t\right) e^{-2in\omega t}|\phi _{\alpha }^{R}\rangle ~,
\label{psi}
\end{equation}%
where $W_{\alpha ,n}\left( t\right) $ are time-dependent expansion
coefficients. Using the standard Floquet formalism described in \ref%
{ap1}, one obtains the system of equations%
\begin{equation}
i\dot{W}_{\beta ,m}\left( t\right) =\sum_{\alpha =\pm }\sum_{n=-\infty
}^{\infty }~_{S}\langle \beta ,m|{H}_{F}^{\left( n\right) }|\phi _{\alpha
},n\rangle _{S}~W_{\alpha ,n}\left( t\right) ~,
\end{equation}%
where ${H}_{F}^{\left( n\right) }$ is the Floquet Hamiltonian, Eq.~(\ref{fl}%
), and $|\phi _{\alpha },n\rangle _{S}\equiv \left\vert \phi _{\alpha
}^{R}\right\rangle \otimes \left\vert n\right) $ is a basis vector in the
enlarged Sambe space.

We evaluate the resonant Floquet matrix elements by Brillouin--Wigner
partitioning \cite{Lowdin1962}. Let $P$ project onto the nearly degenerate
resonant subspace, with eigenenergy $\varepsilon $, and let $Q=\hat{I}-P$
project onto its complement. As shown in \ref{ap2}, in the subspace defined
by the projector $P$, the effective Hamiltonian is 
\begin{equation}
{H}_{P}^{\left( n\right) }=P\left[ {H}_{F}^{\left( n\right)
}+V_{S}Q\sum_{l=0}^{\infty }\left( G_{0}QV_{S}Q\right) ^{l}G_{0}QV_{S}\right]
P\,~,  \label{forma}
\end{equation}%
where%
\begin{equation}
G_{0}\equiv \left( \varepsilon -Q{H}_{F,0}^{\left( n\right) }Q\right) ^{-1}~.
\end{equation}

Near the $\left( 2K+1\right) $-photon resonance, the resonant subspace is 
\begin{equation}
P=|F_{0}\rangle \langle F_{0}|+|F_{K}\rangle \langle F_{K}|\,,
\end{equation}%
where 
\begin{equation}
|F_{0}\rangle =|\phi _{+},0\rangle _{S}~,~|F_{K}\rangle =|\phi
_{-},-K\rangle _{S}\,.
\end{equation}%
The two states are nearly degenerate in quasienergy and are connected only
through a sequence of counter-rotating transitions in the intermediate
Floquet sectors. Because the exact quasienergy, $\varepsilon $, enters the
Brillouin--Wigner resolvent $G_{0}$, the effective Hamiltonian ${H}%
_{P}^{\left( n\right) }$ is formally energy dependent. At the order retained
here, however, we evaluate the resolvent at a
reference unperturbed quasienergy $E_0$ of the nearly degenerate pair. The difference between $%
\varepsilon $ and $E_{0}$ contributes only beyond the perturbative order
kept in the effective couplings and diagonal shifts.

In the ordered resonant basis, we write the matrix representing the
effective Hamiltonian ${H}_{P}^{\left( n\right) }$ as%
\begin{equation}
\hat{M} =\frac{M_{00}+M_{KK}}{2}+\left( 
\begin{array}{cc}
\frac{M_{00}-M_{KK}}{2} & M_{0K} \\ 
M_{K0} & -\frac{M_{00}-M_{KK}}{2}%
\end{array}%
\right) \,, 
\end{equation}%
where $M_{ij}\equiv \langle F_{i}|{H}_{P}^{\left( n\right) }|F_{j}\rangle $
and, for the current problem, $M_{0K}=M_{K0}$. The scalar term in this
matrix expression is understood to multiply the identity operator. 
{The eigenvalues of the effective resonant block provide a direct estimate of
the correction introduced by the Brillouin--Wigner coupling. They are
\begin{equation}
\varepsilon_{K,\pm}
=
\frac{M_{00}+M_{KK}}{2}
\pm
\frac{1}{2}
\sqrt{(M_{00}-M_{KK})^2+4M_{0K}M_{K0}}.
\end{equation}
The $Q$ subspace contains all Floquet states
$|\varphi_\alpha,n\rangle_S$, with $\alpha=\pm$ and $n\in\mathbb{Z}$,
except the resonant pair $(+,0)$ and $(-,-K)$. Their unperturbed
quasienergies are $\lambda_\alpha-2n\omega$. At the perturbative order
retained here, the quasienergy entering the Brillouin--Wigner resolvent may
be evaluated at $M_{00}$ provided that
\begin{equation}
\max_{\sigma=\pm}
\left|\varepsilon_{K,\sigma}-M_{00}\right|
\ll
\min_{(\alpha,n)\notin\{(+,0),(-,-K)\}}
\left|M_{00}-\left(\lambda_\alpha-2n\omega\right)\right|,
\end{equation}
and, analogously, it may be evaluated at $M_{KK}$ provided that
\begin{equation}
\max_{\sigma=\pm}
\left|\varepsilon_{K,\sigma}-M_{KK}\right|
\ll
\min_{(\alpha,n)\notin\{(+,0),(-,-K)\}}
\left|M_{KK}-\left(\lambda_\alpha-2n\omega\right)\right|.
\end{equation}
If both inequalities are satisfied, either diagonal quasienergy may be used
as the reference value in the resolvent at the perturbative order retained
here. Near resonance, where $M_{00}\simeq M_{KK}$ and
$M_{0K}=M_{K0}$, the quantities on the left-hand sides are of the order of
the effective resonant splitting, $|M_{0K}|$. Thus, the relevant condition
is that the effective multiphoton coupling remain small compared with the
quasienergy separation between the resonant pair and the nearest eliminated
Floquet state. }

After straightforward calculations, we find that the leading nonzero
off-diagonal matrix elements for the lowest odd resonances are%
\begin{equation}
M_{01}=-\frac{g\left( \chi +\delta \right) }{2\chi }
\end{equation}%
\begin{equation}
M_{02}=\frac{g^{3}}{8\omega \chi }\frac{\chi +\delta }{\chi -\omega }
\end{equation}%
\begin{equation}
M_{03}=-\frac{g^{5}}{\chi }\frac{\chi +\delta }{2^{6}\omega ^{2}\left( \chi
-2\omega \right) \left( \chi -\omega \right) }\,.
\end{equation}%
The diagonal elements are%
\begin{equation}
M_{00}=\lambda _{+}+\mu _{K}~,~M_{KK}=\lambda _{-}+2K\omega -\mu _{K}~,
\end{equation}%
where%
\begin{equation}
\mu _{K=1}=\frac{g^{2}\left( \chi -\delta \right) ^{2}}{2^{3}\chi ^{2}\left(
\chi +\omega \right) }
\end{equation}%
\begin{equation}
\mu _{K>1}=\frac{g^{2}}{2^{3}\chi ^{2}}\left[ \frac{\left( \chi -\delta
\right) ^{2}}{\chi +\omega }+\frac{\left( \chi +\delta \right) ^{2}}{\chi
-\omega }\right] \,.
\end{equation}

Including the leading diagonal shifts, the $(2K+1)$-photon resonance is
defined by the {\em zero-real-detuning condition} $\func{Re}\left( d\right) =0$,
where%
\begin{equation}
d\equiv \chi +\mu _{K}-K\omega \,.
\end{equation}%
In the non-Hermitian problem, this condition cancels the real part of the
quasienergy mismatch between the two resonant states, but does not imply
equality of their complete complex quasienergies. The remaining imaginary
mismatch $\func{Im}(d)$ is retained explicitly in the effective dynamics.

After imposing the resonance condition and removing an overall real energy
shift that contributes only a global phase, $\hat{M}$ takes the form%
\begin{equation}
\hat{M}=-i\frac{\kappa }{2}+\hat{m}~,~\hat{m}=\left( 
\begin{array}{cc}
i\func{Im}\left( d\right) & M_{0K} \\ 
M_{0K} & -i\func{Im}\left( d\right)%
\end{array}%
\right) \,.  \label{eqm}
\end{equation}%
Denoting $\mathbf{V}=\left( W_{+,0},W_{-,-K}\right) ^{T}$, the projected
dynamics obeys the Schr\"{o}dinger-type equation $i\mathbf{\dot{V}}=\hat{M}%
\mathbf{V}$. Its solution is%
\begin{equation}
\mathbf{V}\left( t\right) =e^{-\kappa t/2}\left( \cos \left( \eta t\right) 
\hat{I}-i\frac{\sin \left( \eta t\right) }{\eta }\hat{m}\right) \mathbf{V}%
\left( 0\right) ~,
\end{equation}%
where%
\begin{equation}
\eta \equiv \sqrt{M_{0K}^{2}-\left[ \func{Im}\left( d\right) \right] ^{2}}~.
\end{equation}%
The corresponding amplitudes are%
\begin{eqnarray}
W_{+,0}\left( t\right) &=&e^{-\kappa t/2}\left[ \left( \cos \left( \eta
t\right) +\frac{\func{Im}\left( d\right) }{\eta }\sin \left( \eta t\right)
\right) W_{+,0}\left( 0\right) \right.  \notag \\
&&-\left. i\frac{M_{0K}}{\eta }\sin \left( \eta t\right) W_{-,-K}\left(
0\right) \right]
\end{eqnarray}%
\begin{eqnarray}
W_{-,-K}\left( t\right) &=&e^{-\kappa t/2}\left[ \left( \cos \left( \eta
t\right) -\frac{\func{Im}\left( d\right) }{\eta }\sin \left( \eta t\right)
\right) W_{-,-K}\left( 0\right) \right.  \notag \\
&&-\left. i\frac{M_{0K}}{\eta }\sin \left( \eta t\right) W_{+,0}\left(
0\right) \right] \,.
\end{eqnarray}%
For the atom initially in $|e\rangle $, the dressed-state amplitudes are%
\begin{equation}
W_{+,0}\left( 0\right) =\frac{\chi -\delta }{\mathcal{N}_{+}}%
~,~W_{-,-K}\left( 0\right) =-\frac{\chi +\delta }{\mathcal{N}_{-}}\,\,.
\end{equation}

\subsection{Conditional transfer time and ideal mean time to successful
preparation}

Transforming from Sambe space back to the physical Hilbert space gives, near
the selected multiphoton resonance,%
\begin{equation}
|\psi \left( t\right) \rangle \approx W_{+,0}\left( t\right) |\phi
_{+}^{R}\rangle +W_{-,-K}\left( t\right) e^{2iK\omega t}|\phi
_{-}^{R}\rangle \,.
\end{equation}%
Equivalently,%
\begin{equation}
|\psi (t)\rangle =C_{g}(t)|g\rangle +C_{e}(t)|e\rangle
\end{equation}%
with%
\begin{equation}
C_{g}\left( t\right) \approx g\left[ \frac{W_{+,0}\left( t\right) }{\mathcal{%
N}_{+}}+\frac{W_{-,-K}\left( t\right) }{\mathcal{N}_{-}}e^{2iK\omega t}%
\right]  \label{prec0}
\end{equation}%
and%
\begin{equation}
C_{e}\left( t\right) \approx \frac{W_{+,0}\left( t\right) }{\mathcal{N}_{+}}%
(\chi -\delta )-\frac{W_{-,-K}\left( t\right) }{\mathcal{N}_{-}}(\chi
+\delta )e^{2iK\omega t}\,.
\end{equation}%
The squared norm of the unnormalized state gives the {\em probability of the
selected no-jump trajectory},%
\begin{equation}
P^{\left( c\right) }(t)=|C_{g}(t)|^{2}+|C_{e}(t)|^{2}~,  \label{prec}
\end{equation}%
whereas normalization by this probability gives the conditional ground- and
excited-state populations,%
\begin{equation}
P_{g}^{\left( c\right) }(t)=\frac{|C_{g}(t)|^{2}}{P^{\left( c\right) }(t)}%
~,~P_{e}^{\left( c\right) }(t)=\frac{|C_{e}(t)|^{2}}{P^{\left( c\right) }(t)}%
~.  \label{prec1}
\end{equation}

To quantify the acceleration of the conditioned dynamics from the initial
state $|\psi _{in}\rangle =|e\rangle $, we define $T_{p}$ as the first time
at which the conditional population of the target state $|\psi _{tar}\rangle
=|g\rangle $ reaches a prescribed value $p$, 
\begin{equation}
T_{p}\left( \lambda _{f}\right) =\inf \left\{ t>0:P_{g}^{(c)}(t)\geq
p\right\} .  \label{t1}
\end{equation}%
{To separate the slow multiphoton dynamics from the fast micromotion, we also
introduce the centered period-averaged population of the target state
\begin{equation}
\overline{P}_{\mathrm{tar}}^{(c)}(t)
=
\frac{1}{T_F}
\int_{t-T_F/2}^{t+T_F/2}
P_{\mathrm{tar}}^{(c)}(t')\,dt',
\qquad
T_F=\frac{\pi}{\omega},
\end{equation}
and define the corresponding coarse-grained first-passage time as
\begin{equation}
\overline{T}_p(\lambda_f)
=
\inf\left\{
t>T_F/2:
\overline{P}_{\mathrm{g}}^{(c)}(t)\ge p
\right\}.
\end{equation}}
The \emph{conditional speedup} relative to the
Hermitian dynamics is then defined as 
\begin{equation}
R_{p}(\lambda _{f})=\frac{T_{p}(0)}{T_{p}(\lambda _{f})}.  \label{t2}
\end{equation}%
This quantity characterizes the time scale of the selected no-jump
trajectories. It does not, however, account for the cost associated with
discarding trajectories in which the monitored jump occurs.

To quantify this cost, we introduce the ideal minimum mean time required to
obtain one successful preparation. We assume an ideal repeated-attempt
protocol in which a monitored jump immediately terminates the current
attempt and the system is instantaneously restored to the same initial
state. Here, a successful attempt denotes a no-jump realization that
survives up to $T_{p}$, at which the conditional target-state population has
reached the prescribed value $p$; no additional projective verification of
the target state is assumed.

Consider $N$ identical attempts of the protocol and let $N_{c}(t)$ be the
number of attempts in which no monitored jump has occurred up to time $t$.
At $t=T_{p}\left( \lambda _{f}\right) $, the surviving attempts are
precisely those accepted as successful realizations of the required
conditioned evolution. Hence, the number of successful attempts is 
\begin{equation}
N_{\mathrm{s}}=N_{c}(T_{p}).  \label{t3}
\end{equation}%
The total time spent in all attempts can be obtained directly by counting
the number of attempts that remain active at each instant. During an
infinitesimal interval $dt$ at time $t$, only the $N_{c}(t)$ attempts that
have not yet undergone the monitored jump contribute to the elapsed
preparation time. Therefore, 
\begin{equation}
T_{\mathrm{tot}}=\int_{0}^{T_{p}}N_{c}(t)\,dt.  \label{t4}
\end{equation}%
This expression automatically accounts for both successful and failed
attempts. An attempt in which a monitored jump occurs at some $t<T_{p}$
contributes only up to the jump time, whereas a successful attempt
contributes throughout the full interval $T_{p}$. The mean total attempt
time per successful preparation is consequently 
\begin{equation}
\tau_p=\frac{T_{\mathrm{tot}}}{N_{\mathrm{s}}}%
=\frac{\int_{0}^{T_{p}}[N_{c}(t)/N]\,dt}{N_{c}(T_{p})/N}.  \label{t5}
\end{equation}

For a sufficiently large ensemble, the fraction of attempts that remain in
the no-jump ensemble up to time $t$ is precisely $P^{\left( c\right) }(t)$,
Eq.~(\ref{prec}). Equation~(\ref{t5}) therefore becomes 
\begin{equation}
\tau_p = \frac{\int_{0}^{T_{p}}P^{\left(
c\right) }(t)\,dt} {P^{\left( c\right) }(T_{p})}.  \label{t6}
\end{equation}%
We refer to $%
\tau_p$ as the {\em ideal minimum mean time to
successful preparation}. It represents an optimistic operational benchmark:
an unsuccessful attempt is assumed to terminate immediately when the
monitored photon is detected, and a new attempt is assumed to begin with no
additional preparation or reset time. Any experimental overhead would
therefore increase the actual mean preparation time. In the Hermitian limit, 
$\lambda _{f}=0$, one recovers $\tau_p=T_{p}$.
For the coarse-grained threshold we analogously define
\begin{equation}
\overline{\tau}_p=\frac{\int_0^{\overline T_p}P^{(c)}(t)\,dt}
{P^{(c)}(\overline T_p)}.
\label{taupavg}
\end{equation}
Only the target population is coarse grained; the no-jump survival
probability in Eqs.~(\ref{t6}) and (\ref{taupavg}) is the raw survival
probability of the physical conditioned trajectory.

\subsection{Comparison to numerical results}

\begin{figure}[tbh]
\begin{center}
\includegraphics[width=0.48\textwidth]{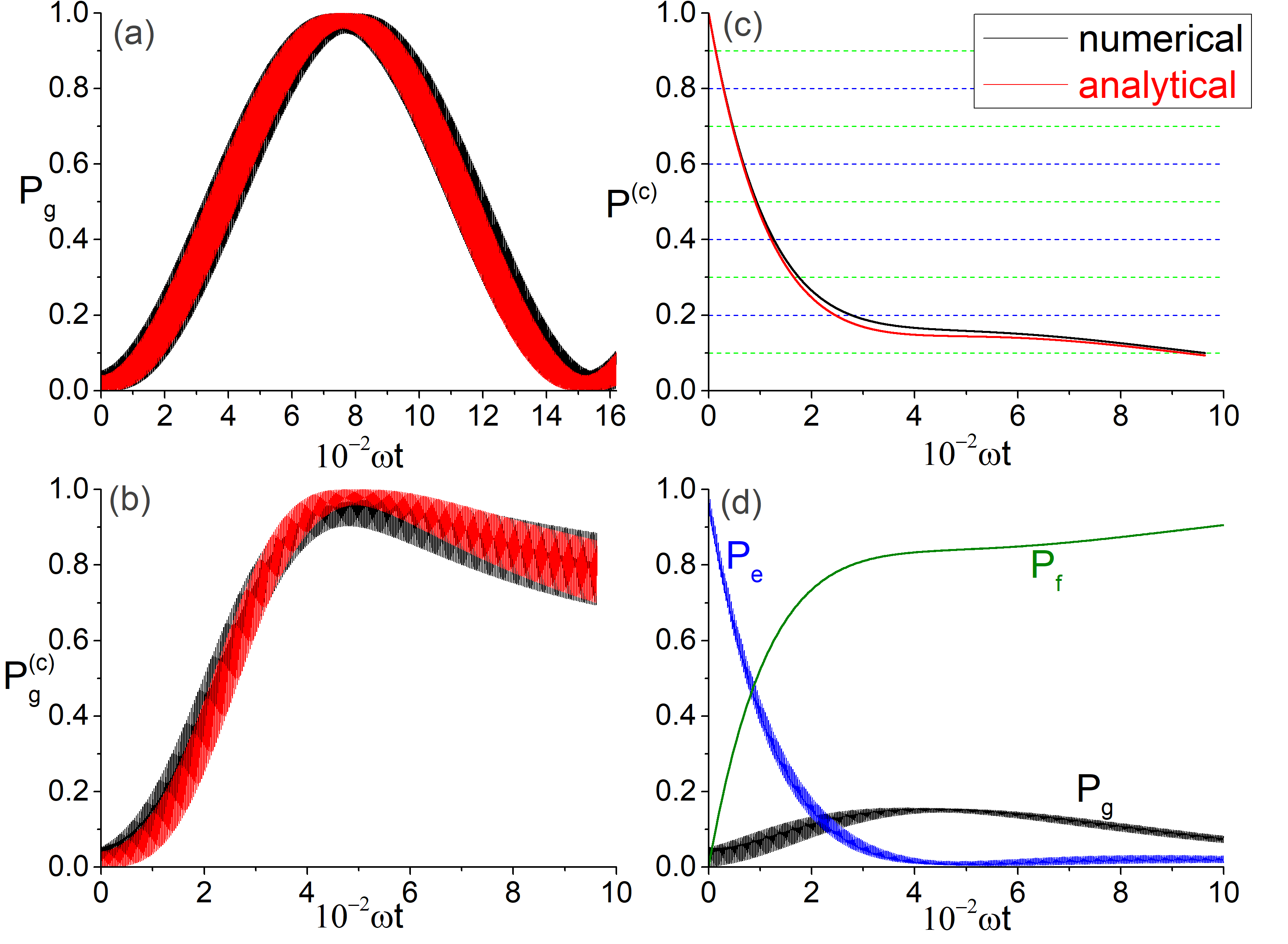} {}
\end{center}
\caption{Semiclassical Rabi model at the three-photon resonance. (a)
Ground-state population for the initial state $|e\rangle $ under unitary
evolution; the black and red curves are the numerical and analytical
results, respectively. (b) Ground-state population in the conditioned
no-jump dynamics. The first analytical and numerical transfer maxima occur
at essentially the same time and earlier than in panel (a). (c) Probability
of no jump during the interval $[0,t)$. (d) Atomic populations in the
unconditioned dissipative evolution generated by the complete master
equation (solved numerically).}
\label{fig2}
\end{figure}

\begin{figure}[tbh]
\begin{center}
\includegraphics[width=0.48\textwidth]{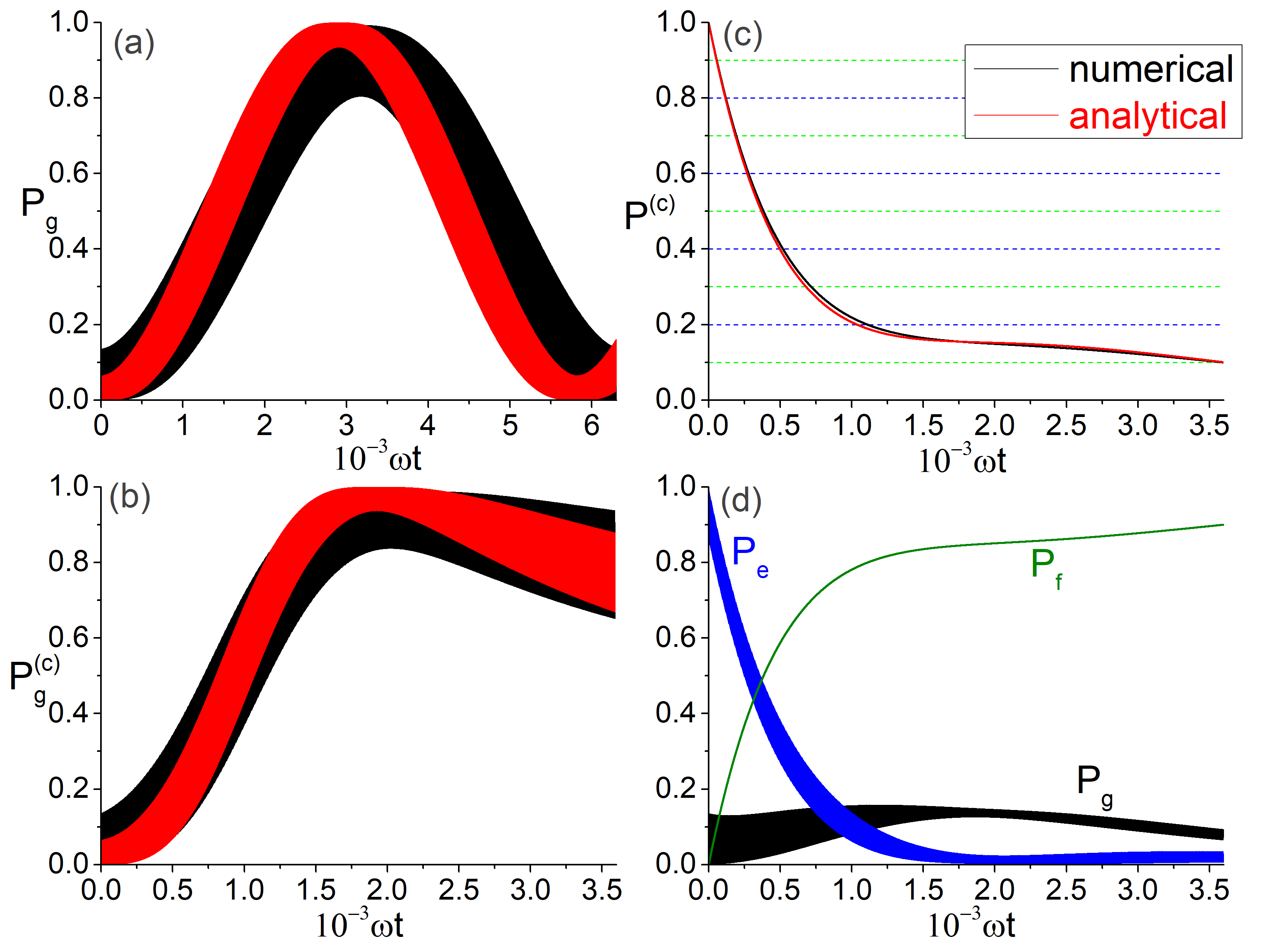} {}
\end{center}
\caption{Same quantities as in Fig. \protect\ref{fig2}, but for the
five-photon resonance. The horizontal axis spans a longer time interval
because the five-photon transition is intrinsically slower, even though a
larger coupling strength is used.}
\label{fig3}
\end{figure}

Figures \ref{fig2} and \ref{fig3} compare the approximate analytical results
with the semiclassical numerical dynamics for the atom initially in $%
|e\rangle $ at the three- and five-photon resonances, respectively. For Fig. %
\ref{fig2}, the parameters are $K=1$, $E_{e}=2.939\omega $, $E_{f}=1.5\omega 
$, $g=0.2\omega $, $g_{f}=0.01\omega $, $\lambda =\gamma _{\phi }=2\times
10^{-4}\omega $, and $\lambda _{f}=8\times 10^{-3}\omega $. For Fig. \ref%
{fig3}, they are $K=2$, $E_{e}=4.78885\omega $, $E_{f}=1.5\omega $, $%
g=0.5\omega $, $g_{f}=0.01\omega $, $\lambda =\gamma _{\phi }=2\times
10^{-5}\omega $, and $\lambda _{f}=2\times 10^{-3}\omega $. The numerical
curves are obtained by integrating the conditioned master equation with the
Hamiltonian (\ref{Hc}) using a fifth--sixth-order Runge--Kutta--Verner
method \footnote{%
In all the simulations throughout this paper, numerical convergence was
checked by reducing the Runge--Kutta time step from $10^{-3}\omega ^{-1} $
to $10^{-4}\omega ^{-1}$, with no discernible changes in the reported
results.}; the remaining curves are the corresponding analytical
approximations. The different horizontal scales make explicit the rapid
growth of the transition time with resonance order: the five-photon process
remains much slower even though the coupling is increased from $0.2\omega $
to $0.5\omega $.

Panels (a) show the ground-state population $P_{g}$ under unitary
evolution. The analytical approximation reproduces the numerical Rabi
oscillations despite treating the auxiliary level $|f\rangle$ only through
its leading dispersive correction. Panels (b) display $P_{g}^{(c)}$ during
the monitored no-jump dynamics. The transition $|e\rangle \rightarrow
|g\rangle $ reaches its first maximum earlier than in the unitary case,
and the analytical and numerical curves predict essentially the same
transfer time. The visible broadening of the curves is caused by rapid
micromotion. Panels (c) show the probability that no $|e\rangle \rightarrow
|f\rangle $ jump has occurred up to time $t$. For the chosen parameters, the
no-jump probability at the first transfer maximum is approximately 15\%.
Panels (d) show the unconditioned evolution obtained by solving numerically the complete
master equation (\ref{cme}), with no postselection of individual
trajectories. The plotted quantities are $P_{e}$, $P_{g}$, and $P_{f}$. In
this ensemble-averaged dynamics, $P_{g}$ remains small, $P_{g}\lesssim 0.2$.
The enhanced population transfer to $|g\rangle $ is therefore a property of
the selected no-jump ensemble rather than of the unconditioned dissipative
evolution. The same mechanism extends to higher odd-multiphoton resonances,
although their characteristic time scales increase rapidly with resonance
order.

\begin{figure}[tbh]
\begin{center}
{\includegraphics[width=0.48\textwidth]{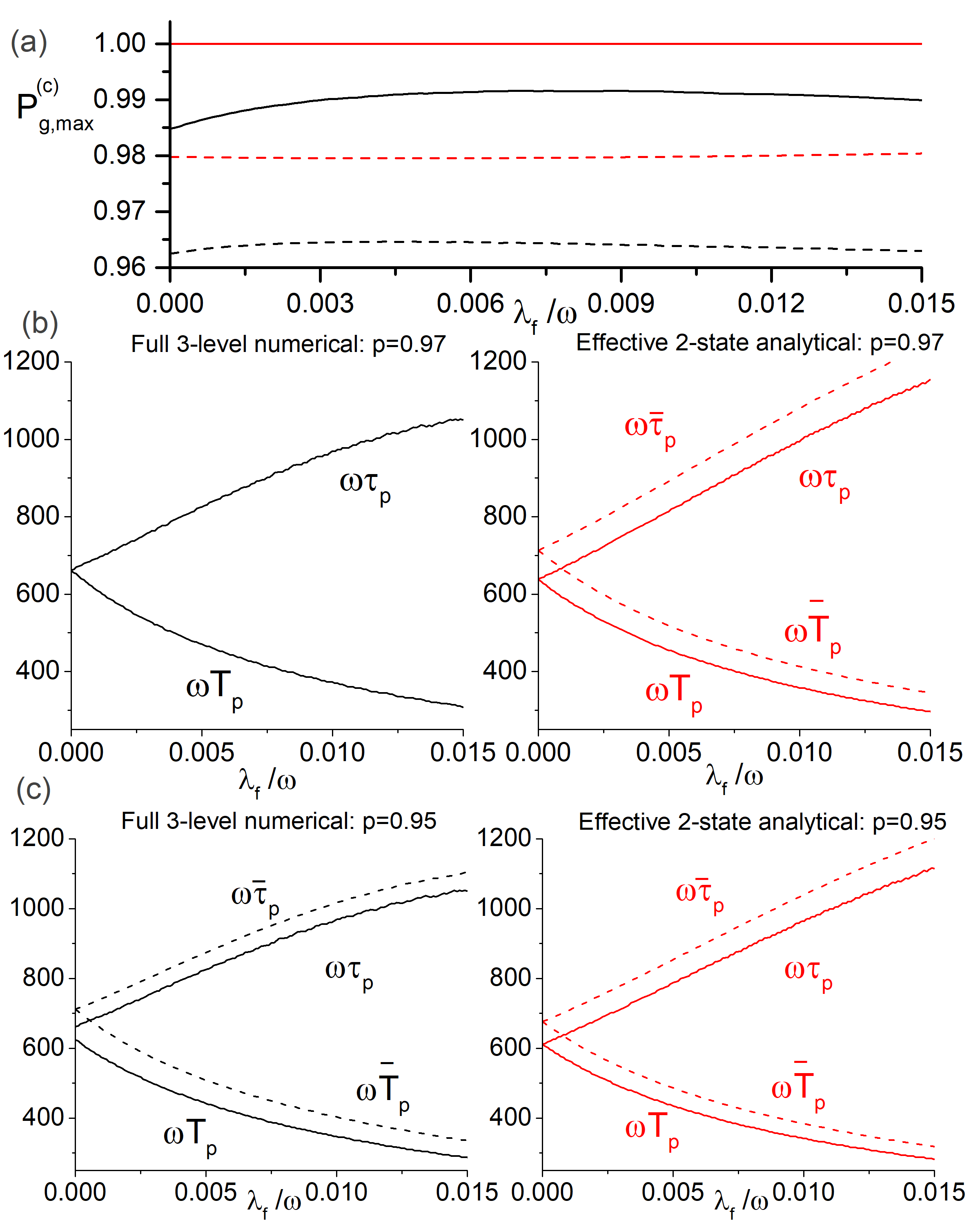}}
\end{center}
\caption{Semiclassical three-photon resonance. (a) Maximum
conditional ground-state population $P_{g,\max}^{(c)}$ versus
$\lambda_f/\omega$, evaluated for $0\leq \omega t\leq 2\times 10^3$. Solid and
dashed curves denote instantaneous and centered period-averaged maxima,
respectively; black denotes the full three-level numerical dynamics and red
the effective two-state analytical result. (b) Transfer times $T_{0.97}$ and
$\overline T_{0.97}$ and corresponding ideal mean preparation times
$\tau_{0.97}$ and $\overline\tau_{0.97}$. Undefined averaged quantities are
not plotted. (c) Same quantities for $p=0.95$. Parameters are $K=1$,
$E_e=2.939\omega$, $E_f=1.5\omega$, $g=0.2\omega$, $g_f=0.01\omega$, and
$\lambda=\gamma_\phi=2\times10^{-5}\omega$.}
\label{fig4}
\end{figure}

{Figure~\ref{fig4} shows the threshold and micromotion behavior
for the three-photon resonance, using $K=1$, $E_e=2.939\omega$,
$E_f=1.5\omega$, $g=0.2\omega$, $g_f=0.01\omega$, and
$\lambda=\gamma_\phi=2\times10^{-5}\omega$. Panel (a) shows the maximum
attainable conditional ground-state population as a function of
$\lambda_f$. Solid curves are instantaneous maxima and dashed curves are
maxima of the centered period-averaged population, with black denoting the
full three-level numerical dynamics and red the effective two-state
analytical theory. For the numerical scans, all maxima and first-passage searches are restricted
to the fixed observation window
\begin{equation}
0\le t\le t_{\max},\qquad t_{\max}=2000/\omega .
\end{equation}
Accordingly,
\begin{equation}
P_{g,\max}^{(c)}(\lambda_f)
=
\max_{0\le t\le t_{\max}}
P_g^{(c)}(t;\lambda_f),
\end{equation}
with the averaged maximum defined analogously from
$\overline P_g^{(c)}$. The finite observation window is chosen to focus on
the resonant multiphoton-transfer dynamics, rather than on the long-time
increase of the ground-state population produced by weak dissipative
relaxation into $|g\rangle$. A threshold is regarded as attainable only if
it is reached within this same time window.

The full three-level numerical result exhibits a small initial increase of
the maximum conditional population at weak monitoring, which is absent from
the effective two-state description. This difference is associated with the
weak unmonitored relaxation and dephasing channels, which are retained in
the complete numerical dynamics but neglected in the analytical
approximation. When both $\lambda$ and $\gamma_\phi$ are set to zero, the
maximum conditional population decreases from the onset of monitoring.
Thus, the small initial nonmonotonicity is a correction arising from the
additional dissipative dynamics rather than an intrinsic feature of the
monitored two-state mechanism. A weaker corresponding variation is also
visible in the period-averaged numerical maximum. At larger $\lambda_f$,
the maximum conditional population decreases.

For this parameter set, we consider the two representative high-fidelity thresholds $p=0.97$ and $p=0.95$, chosen according to the attainable target population and their sensitivity to micromotion. In the full three-level
dynamics, the period-averaged maximum remains below $0.97$ but above $0.95$
throughout the plotted interval. The two thresholds therefore probe
qualitatively different situations: $p=0.97$ is reached by the
instantaneous dynamics but not by its slow period-averaged envelope,
whereas $p=0.95$ remains attainable after coarse graining.

Panel (b) shows the results for $p=0.97$. In the full three-level
calculation, $T_{0.97}$ is well defined and decreases with increasing
$\lambda_f$, while $\tau_{0.97}$ increases. In contrast,
$\overline T_{0.97}$ and $\overline\tau_{0.97}$ are undefined because
$\overline P_{g,\max}^{(c)}<0.97$. The instantaneous crossing of this
threshold is therefore assisted by micromotion. The effective two-state
theory has a larger period-averaged maximum $\overline P_{g,\max}^{(c)}$ and consequently still reaches
$p=0.97$ after coarse graining.

Panel (c) shows the results for $p=0.95$, for which both instantaneous and
period-averaged quantities are defined in the full model. Both $T_{0.95}$
and $\overline T_{0.95}$ decrease with increasing $\lambda_f$ and display
the same qualitative dependence on the monitoring strength. The
conditional acceleration therefore persists in the slow multiphoton
envelope and is not solely a consequence of a micromotion-assisted
threshold crossing. In contrast, $\tau_{0.95}$ and
$\overline\tau_{0.95}$ increase with $\lambda_f$. The results thus
distinguish the conditional acceleration of the multiphoton dynamics from
the absence of an overall preparation-time advantage.
}

\section{Conditioned quantum Rabi model}

\label{sec5}

We now replace the classical drive by a quantized single mode. In the
large-detuning regime, counter-rotating processes generate odd-multiphoton
resonances of the quantum Rabi model \cite%
{MaLaw2015,Garziano2015,marinho2,Hebert}. For the conditioned problem, we
decompose the Hamiltonian as $H_{c}=H_{0}+V$,
where 
\begin{equation}
H_{0}=\left( \omega +\delta _{e}\sigma _{ee}\right) a^{\dagger }a+\frac{%
\Omega }{2}\sigma _{z}-i\kappa \sigma _{ee}+g\left( a\sigma _{+}+a^{\dagger
}\sigma _{-}\right)
\end{equation}%
and 
\begin{equation}
V=g\left( a^{\dagger }\sigma _{+}+a\sigma _{-}\right) .
\end{equation}%
The complex coefficient $\delta
_{e} $ accounts for the photon-number-dependent dispersive correction
induced by the off-resonant auxiliary level $|f\rangle $, as derived in
\ref{ap3}, 
\begin{equation}
\delta _{e}=g_{f}^{2}\sum_{j=\pm }\left( E_{e}-E_{f}+j\omega -i\kappa
\right) ^{-1}.
\end{equation}%
The photon-number-independent contribution is 
\begin{equation}
\delta _{e,0}=\frac{g_{f}^{2}}{E_{e}-E_{f}-\omega -i\kappa }.
\end{equation}%
Its real and imaginary parts are incorporated, up to an irrelevant real
scalar energy shift, through the replacements 
\begin{equation}
\Omega \rightarrow \Omega +\func{Re}\left( \delta _{e,0}\right) ,\qquad
\kappa \rightarrow \kappa -\func{Im}\left( \delta _{e,0}\right) .
\end{equation}%
For notational simplicity, the symbols $\Omega $ and $\kappa $ below denote
these corrected quantities. To the second order in $g_{f}$ retained here, $%
\delta _{e,0}$ and $\delta _{e}$ are evaluated using the uncorrected
parameters in their denominators, since reinserting the dispersive
corrections there would produce only higher-order terms. The total
second-order shift of the bare state $|e,n\rangle $ is therefore $\delta _{e,n}=\delta _{e,0}+n\delta _{e}$,
in agreement with \ref{ap3}.

The right and left eigenstates (dressed states) of the non-Hermitian
Jaynes--Cummings Hamiltonian $H_{0}$, in the $N$-excitation doublet, satisfy 
\begin{equation}
{}_{L}\langle N,\alpha |M,\beta \rangle _{R}=\delta _{N,M}\delta _{\alpha
\beta }
\end{equation}%
and are 
\begin{equation}
|N,\pm \rangle _{R}=\frac{G_{N}|e,N-1\rangle +r_{N,\pm }|g,N\rangle }{%
\mathcal{N}_{N,\pm }}
\end{equation}%
and 
\begin{equation}
{}_{L}\langle N,\pm |=\frac{G_{N}\langle e,N-1|+r_{N,\pm }\langle g,N|}{%
\mathcal{N}_{N,\pm }}.
\end{equation}%
Here the dressed doublets are labeled by positive integers $N$ and $M$, and $%
\alpha ,\beta =\pm $. The relevant quantities are 
\begin{equation}
G_{N}=g\sqrt{N},\qquad r_{N,\pm }=\delta _{N}\pm \chi _{N},\qquad \chi _{N}=%
\sqrt{\delta _{N}^{2}+G_{N}^{2}},
\end{equation}%
where the photon-number-dependent dispersive correction makes the effective
detuning excitation dependent, 
\begin{equation}
\delta _{N}=\frac{\Delta -(N-1)\delta _{e}+i\kappa }{2}.
\end{equation}%
The normalization factors are 
\begin{equation}
\mathcal{N}_{N,\pm }=\pm \sqrt{2\chi _{N}\left( \chi _{N}\pm \delta
_{N}\right) }.
\end{equation}

The corresponding complex eigenvalues are 
\begin{equation}
\lambda _{N,\pm }=N\left( \omega +\frac{\delta _{e}}{2}\right) -\frac{\omega
+\delta _{e}}{2}-\frac{i\kappa }{2}\pm \chi _{N}.
\end{equation}%
For the purposes of this section, it is useful to note that the initial
state $|e,0\rangle$ can be expanded in the first dressed doublet as
\begin{equation}
|e,0\rangle =\frac{g}{\mathcal{N}_{1,+}}|1,+\rangle _{R}+\frac{g}{\mathcal{N}%
_{1,-}}|1,-\rangle _{R}.
\end{equation}

Near a multiphoton resonance, we employ the same projection construction as
in the Floquet problem. Let $P$ denote the projector onto the resonant
dressed-state subspace and $Q=\hat{I}-P$ the projector onto its complement.
To second order in the counter-rotating perturbation $V$, the effective
Hamiltonian is 
\begin{equation}
H_{\mathrm{eff}}=PH_{0}P+PVP+PVQ(E_{0}-QH_{0}Q)^{-1}QVP\,~,
\end{equation}%
where, at the approximation order considered, $E_0$ is a reference
unperturbed energy chosen within the nearly degenerate resonant pair.

For the three-photon resonance originating from $|e,0\rangle $, the
projector is $P=|1,+\rangle _{R}{}_{L}\langle 1,+|~+~|3,-\rangle
_{R}{}_{L}\langle 3,-|\,$, and the relevant effective matrix elements are%
\begin{equation}
\Theta _{3}\equiv ~_{L}\langle 1,+|H_{\mathrm{eff}}|3,-\rangle
_{R}=~_{L}\langle 3,-|H_{\mathrm{eff}}|1,+\rangle _{R}=\frac{g^{2}r_{1,+}%
\sqrt{3!}}{\mathcal{N}_{3,-}\mathcal{N}_{1,+}}
\end{equation}%
\begin{equation}
_{L}\langle 1,\pm |H_{\mathrm{eff}}|1,\pm \rangle _{R}=\lambda _{1,\pm }+\nu
_{1,\pm }
\end{equation}%
\begin{equation}
_{L}\langle 3,-|H_{\mathrm{eff}}|3,-\rangle _{R}=\lambda _{3,-}+\nu _{3,-}
\end{equation}%
\begin{equation}
\nu _{1,+}=-\frac{2r_{1,+}^{2}G_{3}^{2}}{\mathcal{N}_{1,+}^{2}\mathcal{N}%
_{3,+}^{2}}\frac{g^{2}}{2\omega +\delta _{e}-\chi _{1}+\chi _{3}}
\end{equation}%
\begin{equation}
\nu _{1,-}=-\frac{2r_{1,-}^{2}G_{3}^{2}}{\mathcal{N}_{1,-}^{2}\mathcal{N}%
_{3,+}^{2}}\frac{g^{2}}{2\omega +\delta _{e}+\chi _{1}+\chi _{3}}
\end{equation}%
\begin{eqnarray}
\nu _{3,-} &=&\frac{g^{2}}{\mathcal{N}_{3,-}^{2}}\left[ \frac{r_{1,-}^{2}}{%
\mathcal{N}_{1,-}^{2}}\frac{2G_{3}^{2}}{2\omega +\delta _{e}-\chi _{3}+\chi
_{1}}\right.  \notag \\
&&-\frac{G_{5}^{2}}{\mathcal{N}_{5,+}^{2}}\frac{4r_{3,-}^{2}}{2\omega
+\delta _{e}+\chi _{3}+\chi _{5}}  \notag \\
&&-\left. \frac{G_{5}^{2}}{\mathcal{N}_{5,-}^{2}}\frac{4r_{3,-}^{2}}{2\omega
+\delta _{e}+\chi _{3}-\chi _{5}}\right] \,.
\end{eqnarray}%
For $\kappa \ll g\ll 2\omega $ and $\left\vert \delta _{e}\right\vert \ll
\omega $, one obtains $\Theta _{3}\approx \sqrt{3!}g^{3}/2^{2}\omega ^{2}$.
The cubic weak-coupling scaling arises from one explicit counter-rotating
matrix element together with the dressed-state admixtures already contained
in $H_{0}$.

For the five-photon resonance and the initial state $|e,0\rangle $, the
projector is $P=|1,+\rangle _{R}{}_{L}\langle 1,+|~+~|5,-\rangle
_{R}{}_{L}\langle 5,-|\,$, and the corresponding matrix elements are%
\begin{eqnarray}
\Theta _{5} &\equiv &~_{L}\langle 1,+|H_{\mathrm{eff}}|5,-\rangle
_{R}=~_{L}\langle 5,-|H_{\mathrm{eff}}|1,+\rangle _{R}  \notag \\
&=&-\frac{g^{4}r_{1,+}\sqrt{5!}}{\mathcal{N}_{1,+}\mathcal{N}_{5,-}}\left[ 
\frac{r_{3,+}}{\mathcal{N}_{3,+}^{2}}\frac{1}{2\omega +\delta _{e}-\chi
_{1}+\chi _{3}}\right.  \notag \\
&&+\left. \frac{r_{3,-}}{\mathcal{N}_{3,-}^{2}}\frac{1}{2\omega +\delta
_{e}-\chi _{1}-\chi _{3}}\right]
\end{eqnarray}%
\begin{equation}
_{L}\langle 5,-|H_{\mathrm{eff}}|5,-\rangle _{R}=\lambda _{5,-}+\nu _{5,-}
\end{equation}%
\begin{eqnarray}
\nu _{1,+} &=&-\frac{6g^{4}r_{1,+}^{2}}{\mathcal{N}_{1,+}^{2}}\left[ \frac{1%
}{2\omega +\delta _{e}-\chi _{1}+\chi _{3}}\frac{1}{\mathcal{N}_{3,+}^{2}}%
\right.  \notag \\
&&\left. +\frac{1}{2\omega +\delta _{e}-\chi _{1}-\chi _{3}}\frac{1}{%
\mathcal{N}_{3,-}^{2}}\right]
\end{eqnarray}%
\begin{eqnarray}
\nu _{1,-} &=&-\frac{6g^{4}r_{1,-}^{2}}{\mathcal{N}_{1,-}^{2}}\left[ \frac{1%
}{2\omega +\delta _{e}+\chi _{1}+\chi _{3}}\frac{1}{\mathcal{N}_{3,+}^{2}}%
\right.  \notag \\
&&\left. +\frac{1}{2\omega +\delta _{e}+\chi _{1}-\chi _{3}}\frac{1}{%
\mathcal{N}_{3,-}^{2}}\right]
\end{eqnarray}%
\begin{eqnarray}
\hspace{-12mm}\nu _{5,-} &=&\frac{g^{4}}{\mathcal{N}_{5,-}^{2}}\left[ \frac{%
20r_{3,+}^{2}}{\mathcal{N}_{3,+}^{2}{(2\omega +\delta _{e}-\chi _{5}-\chi
_{3})}}\right.  \notag \\
&&+\frac{20r_{3,-}^{2}}{\mathcal{N}_{3,-}^{2}{(2\omega +\delta _{e}-\chi
_{5}+\chi _{3})}}-\frac{42r_{5,-}^{2}}{\mathcal{N}_{7,+}^{2}{(2\omega
+\delta _{e}+\chi _{5}+\chi _{7})}}  \notag \\
&&\left. +\frac{42r_{5,-}^{2}}{\mathcal{N}_{7,-}^{2}{(2\omega +\delta
_{e}+\chi _{5}-\chi _{7})}}\right] .
\end{eqnarray}%
For $\kappa \ll g\ll 2\omega $ and $\left\vert \delta _{e}\right\vert \ll
\omega $, we have $\Theta _{5}\approx \sqrt{5!}g^{5}/2^{6}\omega ^{4}$.
Notice that, although the five-photon coupling scales as the fifth power of $%
g$, it is generated at second order in the perturbation $V$: the dressed
states of $H_{0}$ already contain powers of the rotating interaction, while
the two applications of $V$ connect the excitation manifolds through the
intermediate $N=3$ doublet. Besides, the corrections $\nu _{n,\pm }$ to the
bare eigenvalues $\lambda _{n,\pm }$ scale as $g^{4}/\omega ^{3}$.

Within the resonant subspace $P_{N}=|1,+\rangle _{R}{}_{L}\langle
1,+|~+~|N,-\rangle _{R}{}_{L}\langle N,-|$, the effective Hamiltonian is
represented by the two-dimensional matrix%
\begin{equation}
\hat{M}^{(N)}=d_{N,+}\hat{I}+\left( 
\begin{array}{cc}
d_{N,-} & \Theta _{N} \\ 
\Theta _{N} & -d_{N,-}%
\end{array}%
\right) ,  \label{prev}
\end{equation}%
where%
\begin{equation}
d_{N,\pm }\equiv \frac{\left( \lambda _{1,+}+\nu _{1,+}\right) \pm \left(
\lambda _{N,-}+\nu _{N,-}\right) }{2}~.
\end{equation}%
The corrected $N$-photon resonance is determined by the zero-real-detuning
condition $\func{Re}\left( d_{N,-}\right) =0$. This condition cancels the
real part of the effective quasienergy mismatch between the two resonant
dressed states, but does not imply equality of their complete complex
quasienergies. At resonance, one sets $d_{N,-}\rightarrow i\func{Im}%
(d_{N,-}) $ in Eq.~(\ref{prev}). The traceless part then has the same
algebraic structure as the semiclassical matrix $\hat{M}$, Eq.~(\ref{eqm}).
Consequently, the previously derived no-jump solution carries over after
replacing the semiclassical effective coupling and detuning by their quantum
counterparts.

\section{Quantum multiphoton dynamics and numerical validation}

\label{sec6}

For the initial state $|e,0\rangle $ and the corrected $N$-photon resonance,
the effective two-state Hamiltonian gives the approximate conditioned
solution%
\begin{equation}
|\Psi _{N}(t)\rangle =c_{1}(t)|1,+\rangle _{R}+c_{N}(t)|N,-\rangle
_{R}+B\left( t\right) |1,-\rangle _{R}\,,  \label{PsiN}
\end{equation}%
where $\eta _{N}\equiv \sqrt{\Theta _{N}^{2}-\left[ \func{Im}\left(
d_{N,-}\right) \right] ^{2}}$, and the coefficients are%
\begin{equation}
c_{1}\left( t\right) =\frac{g}{\mathcal{N}_{1,+}}e^{-itd_{N,+}}\left[ \cos
\left( \eta _{N}t\right) +\frac{\func{Im}\left( d_{N,-}\right) }{\eta _{N}}%
\sin \left( \eta _{N}t\right) \right]
\end{equation}%
\begin{equation}
c_{N}\left( t\right) =-i\frac{g}{\mathcal{N}_{1,+}}e^{-itd_{N,+}}\frac{%
\Theta _{N}}{\eta _{N}}\sin \left( \eta _{N}t\right)
\end{equation}%
\begin{equation}
B\left( t\right) =\frac{g}{\mathcal{N}_{1,-}}~e^{-it\left( \lambda
_{1,-}+\nu _{1,-}\right) }~.
\end{equation}%
Because the projected quantum Hamiltonian has the same two-state structure
as its semiclassical counterpart, the conditioned dynamics follows from the
same analytical solution. The conditional transfer time and the
corresponding speedup are then evaluated using the definitions introduced in
Eqs.~(\ref{t1}) and (\ref{t2}).

Expanding Eq. (\ref{PsiN}) in the bare basis gives the following nonzero
populations associated with $|g,n\rangle $ and $|e,n\rangle $:%
\begin{equation}
P_{e,0}^{\left( c\right) }=\frac{P_{e,0}^{\left( un\right) }}{P^{\left(
c\right) }\left( t\right) },\quad P_{e,0}^{\left( un\right) }=\left\vert
c_{1}(t)\frac{g}{\mathcal{N}_{1,+}}+B\left( t\right) \frac{g}{\mathcal{N}%
_{1,-}}\right\vert ^{2}
\end{equation}%
\begin{equation}
P_{g,1}^{\left( c\right) }=\frac{P_{g,1}^{\left( un\right) }}{P^{\left(
c\right) }\left( t\right) },\quad P_{g,1}^{\left( un\right) }=\left\vert
c_{1}(t)\frac{r_{1,+}}{\mathcal{N}_{1,+}}+B\left( t\right) \frac{r_{1,-}}{%
\mathcal{N}_{1,-}}\right\vert ^{2}
\end{equation}%
\begin{equation}
P_{g,N}^{\left( c\right) }=\frac{P_{g,N}^{\left( un\right) }}{P^{\left(
c\right) }\left( t\right) },\quad P_{g,N}^{\left( un\right) }=\left\vert
c_{N}(t)\frac{r_{N,-}}{\mathcal{N}_{N,-}}\right\vert ^{2}
\end{equation}%
\begin{equation}
P_{e,N-1}^{\left( c\right) }=\frac{P_{e,N-1}^{\left( un\right) }}{P^{\left(
c\right) }\left( t\right) },\quad P_{e,N-1}^{\left( un\right) }=\left\vert
c_{N}(t)\frac{G_{N}}{\mathcal{N}_{N,-}}\right\vert ^{2}\,,
\end{equation}%
where the no-jump probability is%
\begin{equation}
P^{\left( c\right) }\left( t\right) =P_{e,0}^{\left( un\right)
}+P_{g,1}^{\left( un\right) }+P_{g,N}^{\left( un\right) }+P_{e,N-1}^{\left(
un\right) }\,.
\end{equation}

{For all quantum transfer times and speedups reported below, the target state
is the joint atom--field state $|g,N\rangle$ and the target fidelity is
therefore $P_{\mathrm{tar}}^{(c)}(t)\equiv P_{g,N}^{(c)}(t)$. 
For the quantum coarse-grained quantities we use the same centered window
$T_{\mathrm{fast}}=\pi/\omega$, which removes the dominant fast $2\omega$
oscillation; here it is a coarse-graining scale rather than a Floquet period
of an explicitly time-periodic Hamiltonian.}

\subsection{Comparison to numerical results}

\begin{figure}[tbh]
\begin{center}
\includegraphics[width=0.48\textwidth]{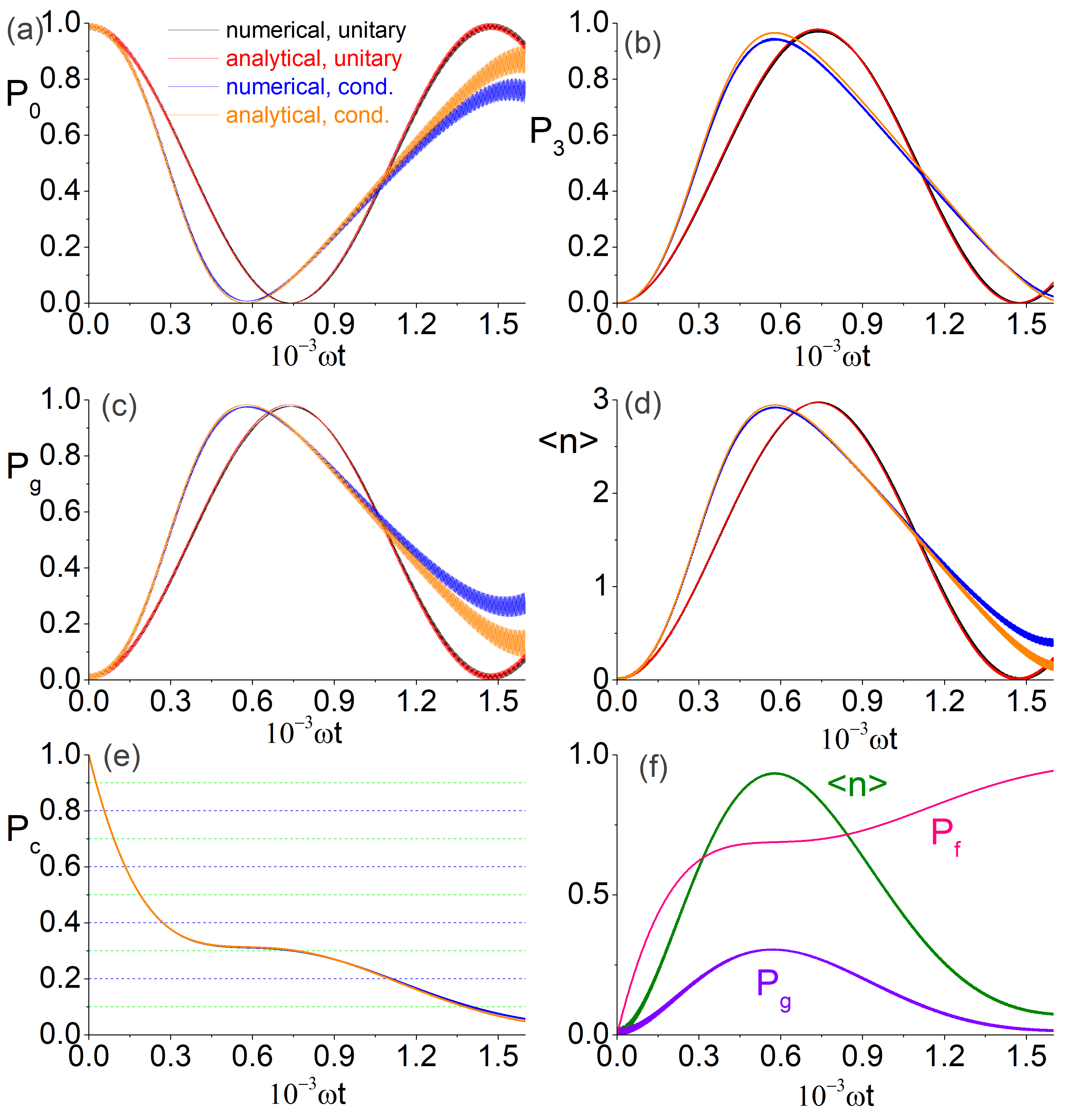} {}
\end{center}
\caption{Quantum Rabi model at the three-photon resonance. (a) Population of
the vacuum Fock state $|0\rangle $: numerical unitary dynamics (black),
analytical unitary dynamics (red), numerical conditioned dynamics (blue),
and analytical conditioned dynamics (orange). (b) Population of the Fock
state $|3\rangle $. (c) Atomic ground-state population $P_g$. (d) Mean
photon number. (e) Probability of no jump during $[0,t)$. (f) Atomic
populations and mean photon number in the unconditioned dissipative
evolution generated by the complete master equation.}
\label{fig5}
\end{figure}

\begin{figure}[tbh]
\begin{center}
\includegraphics[width=0.48\textwidth]{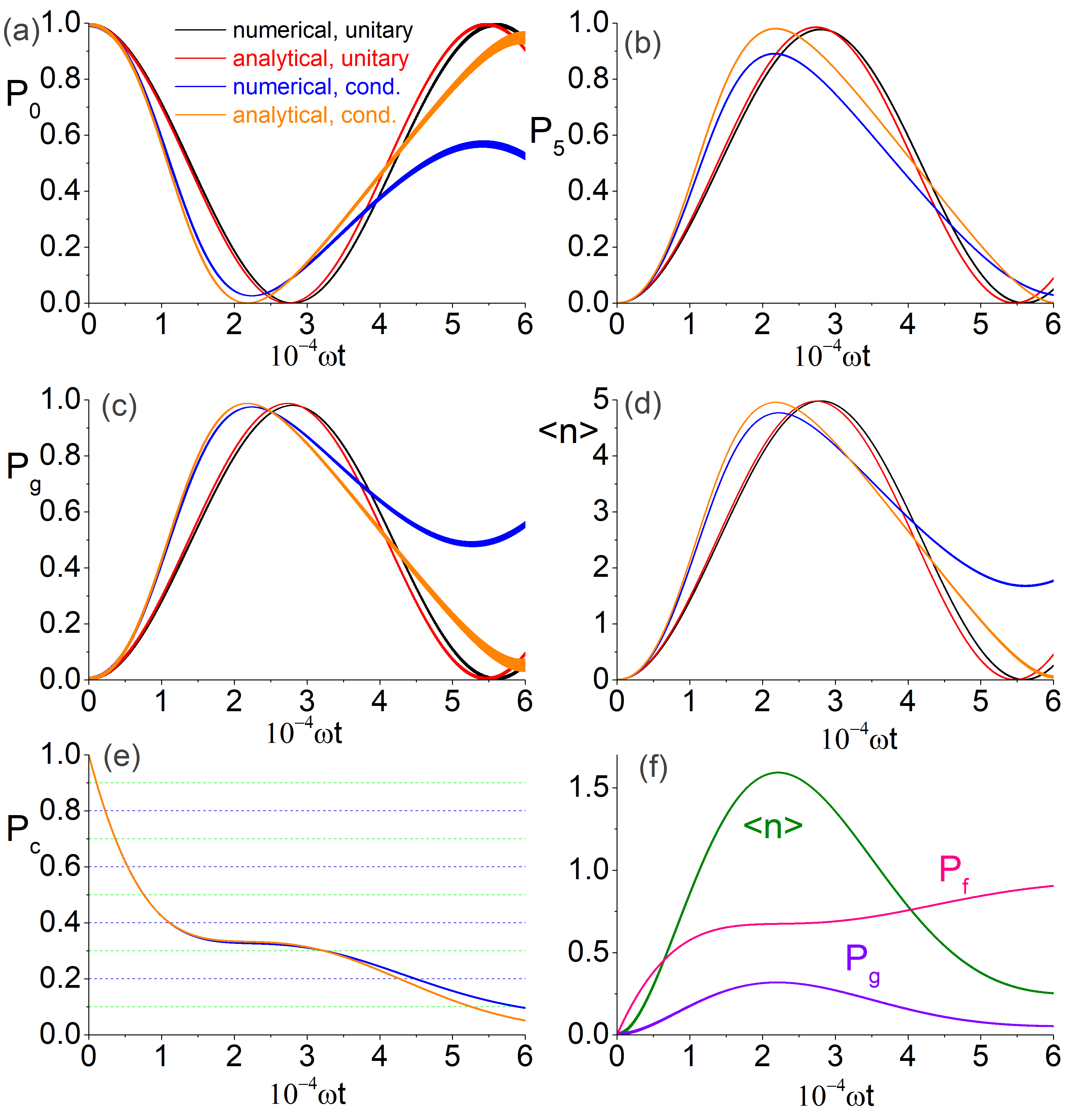} {}
\end{center}
\caption{Same quantities as in Fig. \protect\ref{fig5}, but at the
five-photon resonance. Panel (b) shows the population of the Fock state $%
|5\rangle$.}
\label{fig6}
\end{figure}

Figures \ref{fig5} and \ref{fig6} compare the analytical expressions derived
above with direct numerical integration of the partially conditioned master
equation for the full three-level quantum Rabi model\footnote{%
Convergence with respect to the cavity Hilbert-space truncation was verified
by increasing the maximum photon number from $n_{\max}=9$ to $n_{\max}=11$,
with no discernible changes in the reported results.}. The no-jump
Hamiltonian associated with the monitored atomic decay channel is 
\begin{eqnarray}
H_{c} &=&\omega a^{\dagger }a+E_{e}\sigma _{ee}+E_{f}\sigma _{ff}-i\kappa
\sigma_{ee} \\
&&+g(a+a^{\dagger })\left( \sigma _{eg}+\sigma _{ge}\right)
+g_{f}(a+a^{\dagger })\left( \sigma _{ef}+\sigma _{fe}\right) \,,  \notag
\end{eqnarray}%
whereas the unmonitored cavity-loss channel is retained in Lindblad form
through the additional dissipator $\frac{\lambda _{c}}{2}\mathcal{D}\left(
a\right)\rho _{c}$ in the master equation (\ref{me}), with $\lambda _{c}$
denoting the cavity relaxation rate. Thus, the evolution is conditioned only
on the absence of jumps in the monitored atomic decay channel, while the
cavity decay remains unmonitored.

Figure \ref{fig5} shows the three-photon resonance for $g=0.15\omega $, $%
g_{f}=0.01\omega $, $E_{e}=2.932\omega $, $E_{f}=1.5\omega $, $%
\lambda=\lambda _{c}=\gamma _{\phi }=10^{-5}\omega $, and $\lambda
_{f}=4\times 10^{-3}\omega $. Figure \ref{fig6} shows the five-photon
resonance for $g=0.2\omega $, $g_{f}=0.01\omega $, $E_{e}=4.89949\omega $, $%
E_{f}=3.5\omega $, $\lambda=\lambda _{c}=\gamma _{\phi }=10^{-6}\omega $,
and $\lambda _{f}=10^{-4}\omega $. In both cases, the initial state is $%
|e,0\rangle $. Panels (a) display the probability of the cavity vacuum
state; panels (b) the probabilities of the Fock states $|3\rangle $ and $%
|5\rangle $, respectively; panels (c) the atomic ground-state probability;
panels (d) the average photon number; and panels (e) the no-jump probability.

Throughout the interval in which the no-jump probability exceeds $20\%$, the
analytical curves are in good agreement with the direct numerical solutions,
despite treating the auxiliary level $|f\rangle$ only through its leading
dispersive correction and neglecting the unmonitored dissipative channels in
the analytical treatment. The initial state $|e,0\rangle $ evolves toward
the approximate state $|g,N\rangle $, with $N=3$ and $5$ for the three- and
five-photon resonances, respectively. In both examples, the analytical and
numerical curves reach their first transfer maximum at essentially the same
time, confirming the predicted conditional acceleration. Moreover, the
no-jump probability remains above 30\% throughout the first population
transfer, demonstrating that the speedup does not rely on trajectories with
vanishingly small probability.

Panels (f) show $P_g$, $P_f$, and $\langle n \rangle$ for the unconditioned
evolution, obtained by numerically solving the complete master equation (\ref%
{cme}) with the additional cavity dissipator. These panels confirm that
neither $P_{g}$ nor $\left\langle n\right\rangle $ reaches the corresponding
ideal values $1$ and $N$. The enhancement is therefore specific to the
postselected no-jump ensemble.

\begin{figure}[tbh]
\begin{center}
{\includegraphics[width=0.48\textwidth]{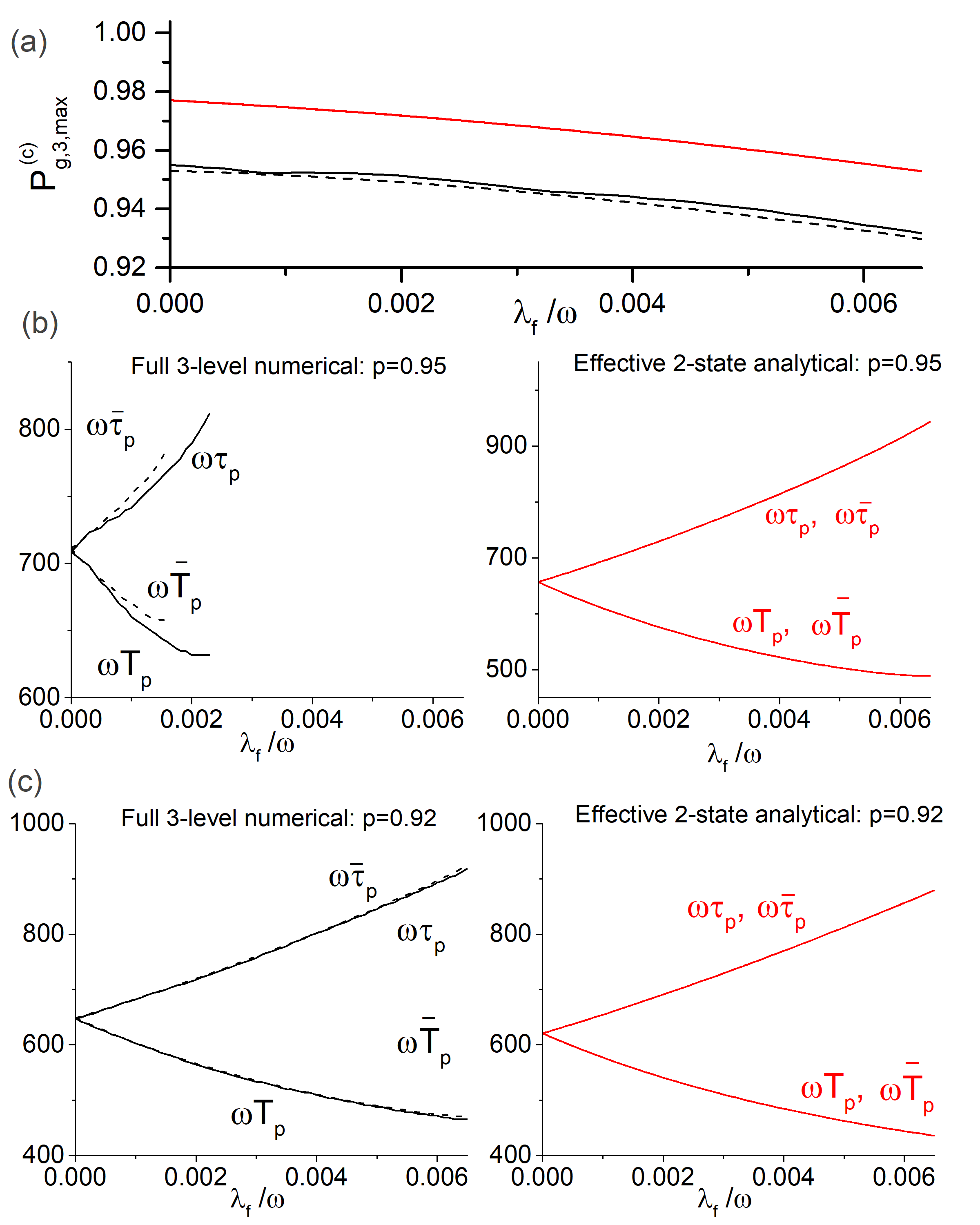}}
\end{center}
\caption{Quantum Rabi model at the three-photon resonance using
the joint target fidelity $P_{g,3}^{(c)}$. (a) Maximum target population within the observation window versus $\lambda_f/\omega$. Black solid and dashed curves are the
instantaneous and coarse-grained full three-level numerical results; the red
curve is the effective two-state analytical result, for which the
instantaneous and coarse-grained maxima are indistinguishable on the plotted
scale. (b) Dimensionless transfer and ideal mean preparation times for
$p=0.95$. The full-model threshold is attainable only over a restricted
monitoring interval, as anticipated from panel (a). (c) Same quantities for
$p=0.92$. Instantaneous and coarse-grained times nearly coincide, while the
transfer times decrease and the mean preparation times increase with
monitoring. }
\label{fig7}
\end{figure}

{Figure~\ref{fig7} shows the conditional target-state dynamics for the quantum
Rabi model at the three-photon resonance, using the joint atom--field
population $P_{g,3}^{(c)}$ as the relevant target fidelity. The parameters
are $E_e=2.932\omega$, $E_f=1.5\omega$, $g=0.15\omega$, and
$g_f=0.01\omega$. To account for realistic dissipative effects, all dissipative channels are retained in the full numerical calculation, with $\lambda=\lambda_c=\gamma_\phi=10^{-5}\omega$. The maximum target
populations are evaluated over the fixed interval
$0\le\omega t\le2\times 10^3$, independently of the chosen threshold.

Panel (a) shows that the maximum target population obtained from the full
three-level calculation is approximately $0.955$ at $\lambda_f=0$ and
decreases as the monitored decay rate is increased. Thresholds above this
value are therefore not attained within the considered time window even in
the absence of monitoring.

For $p=0.95$, panel (b) shows that the numerical transfer time decreases from
approximately $\omega T_p\simeq710$ at $\lambda_f=0$ to about $630$ before
the threshold becomes unattainable, corresponding to a conditional speedup
of approximately $1.13$. Over the same interval, the corresponding ideal
mean preparation time $\tau_p$ increases. The instantaneous and
coarse-grained quantities display the same qualitative dependence on the
monitoring strength.

Panel (c) shows the results for the lower threshold $p=0.92$. In this case,
the instantaneous and coarse-grained transfer times are nearly
indistinguishable throughout the plotted range. This behavior follows
naturally from the effective quantum dynamics. In particular, the target
amplitude $c_N(t)$ in Eq.~(83) is proportional to
$\sin(\eta_N t)$, so that the corresponding target population is governed
primarily by the slow multiphoton oscillation at the effective frequency
$\eta_N$. Consequently, unlike in the semiclassical
Floquet problem, the effective target population contains no explicit fast
micromotion. The small fast components present in the full numerical
dynamics arise from off-resonant contributions beyond the effective
two-state description and therefore produce only a weak difference between
$T_p$ and $\overline T_p$.

In the full numerical calculation, $\omega T_p$ decreases from approximately
$645$ to about $460$, corresponding to a conditional speedup of approximately
$1.4$, while $\omega\tau_p$ increases to about $900$. The effective
two-state description reproduces the same qualitative behavior and gives
very similar instantaneous and coarse-grained transfer times. The close
agreement between $T_p$ and $\overline T_p$ shows that, for this attainable
threshold, the conditional acceleration is associated with the slow
multiphoton dynamics rather than with fast oscillations. At the same time,
the increase of $\tau_p$ and $\overline\tau_p$ shows that the reduction of
the conditional transfer time is not accompanied by a reduction of the mean
preparation time.
}

\section{Conclusions}

\label{sec7}

{We have investigated odd-multiphoton atomic transitions conditioned on
continuous monitoring of the auxiliary decay channel
$|e\rangle\rightarrow|f\rangle$ in a $\Lambda$-type three-level system
coupled to either a classical or a quantized electromagnetic field.
Conditioning on the absence of the monitored emission leads to a
non-Hermitian Rabi-type dynamics. Using Floquet theory and
Brillouin--Wigner projection-operator perturbation theory, we derived
effective two-state Hamiltonians for the relevant multiphoton resonances,
including the high-order couplings generated by counter-rotating virtual
processes and the leading complex dispersive correction induced by the
auxiliary level. The analytical predictions agree well with the full
three-level conditioned dynamics for the three- and five-photon examples
considered.

The conditional acceleration originates from state-selective no-jump
filtering acting on the slow multiphoton dynamics. To distinguish this slow
dynamics from fast oscillatory contributions, we compared instantaneous and
coarse-grained transfer measures and examined the maximum attainable target
population. For the parameter sets considered, the high-fidelity thresholds
used in the numerical examples were chosen according to the attainable
population and are therefore not universal. In the semiclassical model, the
highest threshold considered can be assisted by fast micromotion, whereas
at a slightly lower threshold both the instantaneous and coarse-grained
transfer times decrease with increasing monitoring. This shows that the
conditional acceleration persists at the level of the slow multiphoton
envelope.

In the quantum Rabi model, the relevant target is the joint atom--field
state $|g,N\rangle$, characterized by the conditional population
$P_{g,N}^{(c)}$. The corresponding three-photon results likewise show a
reduction of the transfer time with monitoring, with instantaneous and
coarse-grained quantities remaining very close over the attainable regime.
The effective two-state description reproduces the main features of the
full quantum dynamics, indicating that the observed speedup is associated
with the effective multiphoton transition rather than with fast
oscillations.

The acceleration nevertheless remains conditional on the selected no-jump
ensemble. Although the transfer time can be substantially reduced, the
probability of obtaining the required no-jump trajectory decreases as the
monitoring strength is increased. Consequently, the ideal mean preparation
times $\tau_p$ and $\overline{\tau}_p$ increase in the regimes studied.
Continuous monitoring can therefore accelerate the selected multiphoton
dynamics without providing an overall preparation-time advantage. The main
physical effect is thus a controllable reshaping of the conditional
multiphoton evolution, accompanied by an intrinsic trade-off between faster
transfer within the postselected ensemble and the probability of obtaining
that ensemble.}

\section*{Acknowledgment}

M.V.S.d.P. and A.P.C. acknowledge the financial support by the Brazilian
agency Coordena\c{c}\~{a}o de Aperfei\c{c}oamento de Pessoal de N\'{\i}vel
Superior (CAPES, Finance Code~001).

\appendix{}

\section{Floquet formalism}

\label{ap1}

Substituting $|\psi (t)\rangle $ from Eq. (\ref{psi}) into the Schr\"{o}%
dinger equation gives%
\begin{eqnarray}
&&\sum_{\alpha =\pm }\sum_{n=-\infty }^{\infty }i\dot{W}_{\alpha ,n}\left(
t\right) e^{-2in\omega t}|\phi _{\alpha }^{R}\rangle  \notag \\
&=&\sum_{\alpha =\pm }\sum_{n=-\infty }^{\infty }W_{\alpha ,n}\left(
t\right) \left[ H_{0}-2n\omega +V(t)\right] e^{-2in\omega t}|\phi _{\alpha
}^{R}\rangle ~.
\end{eqnarray}

To convert the periodic problem into a stationary one, we introduce the
auxiliary Fourier Hilbert space $\mathcal{T}$ of square-integrable $T$%
-periodic functions \cite{Sambe1973}. Its scalar product is%
\begin{equation}
(f|g)=\frac{1}{T}\int_{0}^{T}f^{\ast }(t)g(t)dt~.
\end{equation}%
An orthonormal basis of $\mathcal{T}$ is formed by the Fourier states $%
\left\vert n\right) $, with time representation%
\begin{equation}
(t|n)=\exp (-2in\omega t),\qquad n\in \mathbf{Z~}.
\end{equation}%
With this scalar product, the Fourier states obey the orthonormality and
completeness relations%
\begin{equation}
(n|m)=\delta _{nm}~,~\hat{I}_{\mathcal{T}}=\sum_{n=-\infty }^{\infty
}\left\vert n\right) \left( n\right\vert ~.
\end{equation}

The enlarged Floquet, or Sambe, space is the tensor product $\mathcal{H}_{S}=%
\mathcal{H}\otimes \mathcal{T}$, where $\mathcal{H}$ is the physical Hilbert
space. A basis vector in the enlarged space is%
\begin{equation}
|\phi _{\alpha },n\rangle _{S}\equiv \left\vert \phi _{\alpha
}^{R}\right\rangle \otimes \left\vert n\right) ~,
\end{equation}%
where $|\phi _{\alpha }^R\rangle $ is a physical state and $\left\vert
n\right) $ is a Fourier state. Thus,%
\begin{eqnarray}
(t|e^{2i\omega t}|n) &=&e^{2i\omega t}(t|n)=e^{2i\omega t}\exp (-2in\omega t)
\notag \\
&=&\exp \left[ -2i\left( n-1\right) \omega t\right] =(t|n-1)
\end{eqnarray}%
and the Fourier exponentials act as shift operators: $e^{2i\omega t}=S_{-}$
and $e^{-2i\omega t}=S_{+}$, with $S_{\pm }|n)=|n\pm 1)$. The perturbation
in Sambe space is therefore%
\begin{equation}
V_{S}=g\left( S_{-}\sigma _{+}+S_{+}\sigma _{-}\right) \,.
\end{equation}

The Schr\"{o}dinger equation is consequently represented as%
\begin{equation}
\sum_{\beta =\pm }\sum_{m=-\infty }^{\infty }i\dot{W}_{\beta ,m}\left(
t\right) |\phi _{\beta },m\rangle _{S}=\sum_{\alpha =\pm }\sum_{n=-\infty
}^{\infty }W_{\alpha ,n}\left( t\right) {H}_{F}^{\left( n\right) }|\phi
_{\alpha },n\rangle _{S}~,
\end{equation}%
where%
\begin{equation}
{H}_{F}^{\left( n\right) }={H}_{F,0}^{\left( n\right) }+V_{S}~,~{H}%
_{F,0}^{\left( n\right) }=H_{0}-2n\omega \,  \label{fl}
\end{equation}%
are the Floquet Hamiltonians. Projection from the left with $_{S}\langle
\phi _{\beta },m|\equiv \langle \phi _{\beta }^{L}|\otimes (m|$ and use of
biorthogonality yield%
\begin{equation}
i\dot{W}_{\beta ,m}\left( t\right) =\sum_{\alpha =\pm }\sum_{n=-\infty
}^{\infty }~_{S}\langle \beta ,m|{H}_{F}^{\left( n\right) }|\phi _{\alpha
},n\rangle _{S}~W_{\alpha ,n}\left( t\right) ~.
\end{equation}

\section{Brillouin--Wigner perturbation theory}

\label{ap2}

The Floquet eigenvalue problem for ${H}_{F}$ is%
\begin{equation}
{H}_{F}^{\left( n\right) }|\Psi \rangle =\varepsilon |\Psi \rangle 
\end{equation}%
with the decomposition%
\begin{equation}
|\Psi \rangle =|\Psi _{P}\rangle +|\Psi _{Q}\rangle \,,
\end{equation}%
where $\varepsilon $ is the quasienergy, $|\Psi _{P}\rangle =P|\Psi \rangle $
and $|\Psi _{Q}\rangle =Q|\Psi \rangle $. Projection onto the two
complementary subspaces gives%
\begin{equation}
P\mathcal{H}_{F}^{\left( n\right) }P|\Psi _{P}\rangle +P\mathcal{H}%
_{F}^{\left( n\right) }Q|\Psi _{Q}\rangle =\varepsilon |\Psi _{P}\rangle 
\label{e1}
\end{equation}%
\begin{equation}
Q\mathcal{H}_{F}^{\left( n\right) }Q|\Psi _{Q}\rangle +Q\mathcal{H}%
_{F}^{\left( n\right) }P|\Psi _{P}\rangle =\varepsilon |\Psi _{Q}\rangle \,.
\label{e2}
\end{equation}%
Solving Eq. (\ref{e2}) for the nonresonant component gives%
\begin{equation}
|\Psi _{Q}\rangle =\left( \varepsilon -Q\mathcal{H}_{F}^{\left( n\right)
}Q\right) ^{-1}Q\mathcal{H}_{F}^{\left( n\right) }P|\Psi _{P}\rangle \,.
\end{equation}%
Substitution into Eq. (\ref{e1}) produces a closed, energy-dependent
eigenvalue problem in the resonant subspace,%
\begin{equation}
{H}_{P}^{\left( n\right) }|\Psi _{P}\rangle =\varepsilon |\Psi _{P}\rangle
\,,
\end{equation}%
where the Brillouin--Wigner \emph{effective Hamiltonian} in the subspace $P$
is%
\begin{equation}
{H}_{P}^{\left( n\right) }=P\left[ {H}_{F}^{\left( n\right) }+{H}%
_{F}^{\left( n\right) }Q\left( \varepsilon -Q{H}_{F}^{\left( n\right)
}Q\right) ^{-1}Q{H}_{F}^{\left( n\right) }\right] P\,.
\end{equation}

Introducing the unperturbed projected resolvent%
\begin{equation}
G_{0}\equiv \left( \varepsilon -Q{H}_{F,0}^{\left( n\right) }Q\right) ^{-1}~,
\end{equation}%
the full projected resolvent admits the expansion%
\begin{equation}
\left( \varepsilon -Q{H}_{F}^{\left( n\right) }Q\right) ^{-1}=\left(
1-G_{0}QV_{S}Q\right) ^{-1}G_{0}=\sum_{l=0}^{\infty }\left(
G_{0}QV_{S}Q\right) ^{l}G_{0}
\end{equation}%
and the effective Hamiltonian becomes%
\begin{equation}
{H}_{P}^{\left( n\right) }=P\left[ \mathcal{H}_{F}^{\left( n\right)
}+V_{S}Q\sum_{l=0}^{\infty }\left( G_{0}QV_{S}Q\right) ^{l}G_{0}QV_{S}\right]
P\,.  \label{isaq}
\end{equation}%
For each multiphoton resonance, we retain the lowest-order nonvanishing
contribution that couples the selected Floquet states, together with the
leading diagonal corrections required to determine the resonance condition.
This truncation is consistent with the overall level of approximation
adopted in the analytical treatment.

{The Brillouin--Wigner expansion is perturbative when the couplings induced
by $V_S$ between the resonant subspace and the eliminated Floquet states are
small compared with the corresponding quasienergy separations. Since $G_0$
contains the inverse of these separations, a convenient sufficient condition
is
\begin{equation}
\|G_0QV_SQ\|\ll1.
\end{equation}
The same quasienergy gaps determine whether the resolvent may be evaluated
at a reference quasienergy of the resonant pair. Thus, the relevant small
parameters are ratios of coupling matrix elements to quasienergy
differences. Since only quasienergy differences enter these conditions, they are invariant under a global shift of the energy origin.
}

\section{Perturbative elimination of level $|f\rangle $}

\label{ap3}

In the semiclassical regime, the weakly coupled auxiliary level $|f\rangle $
can be eliminated perturbatively. We use Eq.~(\ref{isaq}), set $g=0$, and
choose the projector $P=|e,0\rangle \langle e,0|$. The Fourier components of
the interaction $2g_{f}\cos (\omega t)(\sigma _{ef}+\sigma _{fe})$ couple $%
|e,0\rangle $ to the virtual states $|f,\pm 1\rangle $ with matrix element $%
g_{f}$, where, for this elimination, we define $(t|n)=\exp (-in\omega t)$.
To second order in $g_{f}$, the energy of level $|e\rangle $ is therefore
shifted as $E_{e}\rightarrow E_{e}+\delta _{e}$, where 
\begin{equation}
\delta _{e}=g_{f}^{2}\sum_{j=\pm }\frac{1}{E_{e}-E_{f}+j\omega -i\kappa }.
\label{disp}
\end{equation}%
The perturbative elimination is valid provided $g_{f}\ll \left\vert
E_{e}-E_{f}\pm \omega \right\vert $. In the regime considered here, where $%
\kappa \ll \left\vert E_{e}-E_{f}\pm \omega \right\vert $, the real and
imaginary parts of the shift are approximately 
\begin{equation}
\func{Re}(\delta _{e})\approx g_{f}^{2}\sum_{j=\pm }\left(
E_{e}-E_{f}+j\omega \right) ^{-1},
\end{equation}%
\begin{equation}
\func{Im}(\delta _{e})\approx \kappa g_{f}^{2}\sum_{j=\pm }\left(
E_{e}-E_{f}+j\omega \right) ^{-2}\ll \kappa .
\end{equation}%
Thus, the real part produces the usual dispersive correction to the
transition frequency, whereas the positive imaginary part slightly modifies
the effective decay associated with the term $-i\kappa \sigma _{ee}$. Within
the effective two-level description, these corrections can be included
through 
\begin{equation}
E_{e}\;\longrightarrow \;E_{e}+\func{Re}\left( \delta _{e}\right)
\end{equation}%
and 
\begin{equation}
\kappa \;\longrightarrow \;\kappa -\func{Im}\left( \delta _{e}\right) .
\end{equation}

In the quantum regime, the state $|e,n\rangle$ is coupled by the interaction 
$g_{f}(a+a^{\dagger})(\sigma_{ef}+\sigma_{fe})$ to the virtual states $%
|f,n-1\rangle$ and $|f,n+1\rangle$. Accordingly, we choose 
\begin{equation}
P=|e,n\rangle\langle e,n|
\end{equation}%
and 
\begin{equation}
Q=|f,n-1\rangle\langle f,n-1| +|f,n+1\rangle\langle f,n+1|,
\end{equation}%
with the first term absent for $n=0$. The perturbative elimination is valid
provided $g_{f}\sqrt{n+1}\ll \left|E_{e}-E_{f}-\omega-i\kappa\right|$. The
bare state $|e,n\rangle$ then acquires the second-order complex shift 
\begin{equation}
\delta_{e,n} = \frac{g_{f}^{2}} {E_{e}-E_{f}-\omega-i\kappa} +n\delta_{e},
\end{equation}%
where $\delta_{e}$ is the same complex shift as in the semiclassical regime,
Eq.~(\ref{disp}). For $\kappa\ll\left|E_{e}-E_{f}\pm\omega\right|$, one
obtains 
\begin{equation}
\func{Re}(\delta_{e,n})\approx g_{f}^{2}\left[ \frac{1}{E_{e}-E_{f}-\omega}
+n\sum_{j=\pm} \left(E_{e}-E_{f}+j\omega\right)^{-1} \right],
\end{equation}%
and 
\begin{equation}
\func{Im}(\delta_{e,n})\approx \kappa g_{f}^{2}\left[ \frac{1}{%
(E_{e}-E_{f}-\omega)^{2}} +n\sum_{j=\pm}
\left(E_{e}-E_{f}+j\omega\right)^{-2} \right].
\end{equation}

Equivalently, these corrections can be incorporated into the quantum Rabi
Hamiltonian through the second-order operator 
\begin{equation}
H_{f}^{(2)}=\left[ \frac{g_{f}^{2}}{E_{e}-E_{f}-\omega -i\kappa }+\delta
_{e}a^{\dagger }a\right] \sigma _{ee}.
\end{equation}%
Acting on $|e,n\rangle $, this operator yields precisely the shift $\delta
_{e,n}$ above.

\end{document}